\documentclass[twocolumn,preprint]{aastex63}

\usepackage{amsmath,amstext} 
\usepackage[T1]{fontenc}
\usepackage{apjfonts} 
\usepackage{natbib}
\usepackage{microtype,longtable,verbatim,float,booktabs,graphicx}

\def\etal{et al.}
\newcommand{\ninetychi}{$\chi_{\nu}^{2} (90\%)$}
\newcommand{\target}{RM160}
\newcommand{\targetLong}{J141041.25+531849.0}
\newcommand{\angstrom}{\mbox{\normalfont\AA}}
\newcommand{\MgII}{\hbox{{\rm Mg}\kern 0.1em{\sc II}}}
\newcommand{\CIV}{\hbox{{\rm C}\kern 0.1em{\sc IV}}}
\newcommand{\Ha}{\hbox{{\rm H}$\alpha$}}
\newcommand{\HavNII}{\hbox{({\rm H}$\alpha$+{\rm [N}\kern 0.1em{\sc II}{\rm ]})}}
\newcommand{\Hb}{\hbox{{\rm H}$\beta$}}
\newcommand{\SII}{\hbox{{\rm [S}\kern 0.1em{\sc ii}{\rm ]}}}
\newcommand{\NII}{\hbox{{\rm [N}\kern 0.1em{\sc ii}{\rm ]}}}
\newcommand{\OII}{\hbox{{\rm [O}\kern 0.1em{\sc II}{\rm ]}}}
\newcommand{\OIII}{\hbox{{\rm [O}\kern 0.1em{\sc iii}{\rm ]}}}
\newcommand{\NeIII}{\hbox{{\rm [Ne}\kern 0.1em{\sc III}{\rm ]}}} \newcommand{\NeIIIvHeI}{\hbox{{\rm [Ne}\kern 0.1em{\sc III}{\rm ]}+{\rm [He}\kern 0.1em{\sc I}{\rm ]}}}
\newcommand{\NeV}{\hbox{{\rm [Ne}\kern 0.1em{\sc v}{\rm ]}}}
\newcommand{\HeII}{\hbox{{\rm He}\kern 0.1em{\sc II}}}
\newcommand{\HII}{\hbox{{\rm H}\kern 0.1em{\sc II}}}

\def\arrvline{\hfil\kern\arraycolsep\vline\kern-\arraycolsep\hfilneg}

\newcommand{\LumCont}{$\lambda L_{\lambda} \rm (5100\angstrom)$}
\newcommand{\targetLum}{$4.7 \times 10^{44}$~erg~s$^{-1}$}
\newcommand{\reducedChi}{$\chi_{\nu}^{2} = \frac{1}{\nu} \sum {\frac{\left(O_{i} - M_{i}\right)^{2}}{U_{i}^{2}}}$}

\newcommand{\GrierHbLagObs}{$\tau_{\rm{H\beta,} \rm{obs}} = 31.3^{+8.1}_{-4.1}$ days}
\newcommand{\GrierHaLagObs}{$\tau_{\rm{H\alpha,} \rm{obs}} = 27.7^{+5.3}_{-4.7}$ days}
\newcommand{\HomayouniMgIILagObs}{$\tau_{\rm{MgII,} \rm{obs}} = 144.7^{+24.7}_{-22.6}$ days}

\newcommand{\targetBHMass}{$(M_{\rm BH}/10^{7}M_\odot) = 7.0^{+1.7}_{-1.3}$}

\newcommand{\cosmo}{$\Lambda$CDM cosmology with $\Omega_{\Lambda}$ = 0.7, $\Omega_{M}$ = 0.3, and $H_{0}$ = 70~km~s$^{-1}$~Mpc$^{-1}$}

\shorttitle{Unusual BLR Variability}
\shortauthors{Fries et al.}

\begin{document}

\title{\large \bf The SDSS-V Black Hole Mapper Reverberation Mapping Project: Unusual Broad-Line Variability in a Luminous Quasar}

\author[0000-0001-8032-2971]{Logan B. Fries}
\affil{Department of Physics, 196A Auditorium Road, Unit 3046, University of Connecticut, Storrs, CT 06269, USA}

\author[0000-0002-1410-0470]{Jonathan R. Trump}
\affil{Department of Physics, 196A Auditorium Road, Unit 3046, University of Connecticut, Storrs, CT 06269, USA}

\author[0000-0001-9776-9227]{Megan C. Davis}
\affil{Department of Physics, 196A Auditorium Road, Unit 3046, University of Connecticut, Storrs, CT 06269, USA}

\author[0000-0001-9920-6057]{C.~J.~Grier}
\affiliation{Steward Observatory, The University of Arizona, 933 North Cherry Avenue, Tucson, AZ 85721, USA} 
\affiliation{Department of Astronomy, University of Wisconsin-Madison, Madison, WI 53706, USA} 

\author[0000-0003-1659-7035]{Yue Shen}
\affiliation{Department of Astronomy, University of Illinois at Urbana-Champaign, Urbana, IL 61801, USA}
\affiliation{National Center for Supercomputing Applications, University of Illinois at Urbana-Champaign, Urbana, IL 61801, USA}

\author[0000-0002-6404-9562]{Scott F. Anderson}
\affiliation{Astronomy Department, University of Washington, Box 351580, Seattle, WA 98195, USA}

\author[0000-0002-4459-9233]{Tom Dwelly}
\affiliation{Max-Planck-Institut f{\"u}r extraterrestrische Physik, Giessenbachstra\ss{}e, 85748 Garching, Germany}

\author[0000-0002-3719-940X]{Michael Eracleous}
\affiliation{Department of Astronomy \& Astrophysics and Institute for Gravitation and the Cosmos, The Pennsylvania State University, 525 Davey Lab, University Park, PA 16802, USA}

\author[0000-0002-0957-7151]{Y. Homayouni}
\affiliation{Department of Astronomy \& Astrophysics and Institute for Gravitation and the Cosmos, The Pennsylvania State University, 525 Davey Lab, University Park, PA 16802, USA}

\author[0000-0003-1728-0304]{Keith Horne}
\affiliation{SUPA Physics and Astronomy, University of St. Andrews, Fife, KY16 9SS, Scotland, UK}

\author{Mirko Krumpe}
\affiliation{Leibniz-Institut f\"ur Astrophysik (AIP), An der Sternwarte 16, 14482 Potsdam, Germany}

\author[0000-0002-6770-2627]{Sean Morrison}
\affiliation{Department of Astronomy, University of Illinois at Urbana-Champaign, Urbana, IL 61801, USA}

\author[0000-0001-8557-2822]{Jessie C. Runnoe}
\affiliation{Department of Physics and Astronomy, Vanderbilt University, Nashville, TN 37235, USA}

\author[0000-0002-3683-7297]{Benny Trakhtenbrot}
\affiliation{School of Physics and Astronomy, Tel Aviv University, Tel Aviv 69978, Israel}

% Project 77 (alphabetically)

\author[0000-0002-9508-3667]{Roberto J. Assef}
\affiliation{Instituto de Estudios Astrof\'isicos, Facultad de Ingenier\'ia y Ciencias, Universidad Diego Portales, Av. Ej\'ercito Libertador 441, Santiago, Chile}

\author[0000-0002-0167-2453]{W. N. Brandt}
\affiliation{Department of Astronomy \& Astrophysics, 525 Davey Lab, The Pennsylvania State University, University Park, PA 16802, USA}
\affiliation{Institute for Gravitation and the Cosmos, The Pennsylvania State University, University Park, PA 16802, USA}
\affiliation{Department of Physics, 104 Davey Lab, The Pennsylvania State University, University Park, PA 16802, USA}

\author[0000-0002-8725-1069]{Joel Brownstein}
\affiliation{Department of Physics and Astronomy, University of Utah, 115 S. 1400 E., Salt Lake City, UT 84112, USA}

\author[0000-0001-7306-1830]{Collin Dabbieri}
\affiliation{Department of Physics and Astronomy, Vanderbilt University, Nashville, TN 37235, USA}

\author[0000-0003-4444-0115]{Alexander Fix}
\affiliation{Department of Astrophysical \& Planetary Sciences, University of Colorado, 2000 Colorado Ave, Boulder, CO 80309, USA}

\author[0000-0003-0042-6936]{Gloria Fonseca Alvarez}
\affiliation{Department of Physics, 196A Auditorium Road, Unit 3046, University of Connecticut, Storrs, CT 06269, USA}
\affiliation{NSF's National Optical-Infrared Astronomy Research Laboratory, 950 N. Cherry Ave., Tucson, AZ 85719, USA}

\author[0000-0001-9676-730X]{Sara Frederick}
\affiliation{Department of Physics and Astronomy, Vanderbilt University, Nashville, TN 37235, USA}

\author[0000-0002-1763-5825]{P. B. Hall}
\affiliation{Department of Physics and Astronomy, York University, Toronto, ON M3J 1P3, Canada}

\author[0000-0002-6610-2048]{Anton M. Koekemoer}
\affiliation{Space Telescope Science Institute, 3700 San Martin Dr., Baltimore, MD 21218, USA}

\author[0000-0002-0311-2812]{Jennifer I-Hsiu Li}
\affiliation{Department of Astronomy, University of Michigan, Ann Arbor, MI, 48109, USA}

\author[0000-0003-0049-5210]{Xin Liu}
\affiliation{Department of Astronomy, University of Illinois at Urbana-Champaign, Urbana, IL 61801, USA}
\affiliation{National Center for Supercomputing Applications, University of Illinois at Urbana-Champaign, Urbana, IL 61801, USA}

\author[0000-0002-7843-7689]{Mary Loli Martínez-Aldama}
\affiliation{Instituto de F\'isica y Astronom\'ia, Facultad de Ciencias, Universidad de Valpara\'iso, Gran Breta\~na 1111, Valparaíso, Chile}
\affiliation{Departamento de Astronom\'ia, Universidad de Chile, Casilla 36D, Santiago, Chile}

\author[0000-0001-5231-2645]{Claudio Ricci}
\affiliation{Instituto de Estudios Astrof\'isicos, Facultad de Ingenier\'ia y Ciencias, Universidad Diego Portales, Av. Ej\'ercito Libertador 441, Santiago, Chile}
\affiliation{Kavli Institute for Astronomy and Astrophysics, Peking University, Beijing 100871, People's Republic of China}

\author[0000-0001-7240-7449]{Donald P.\ Schneider} 
\affiliation{Department of Astronomy \& Astrophysics and Institute for Gravitation and the Cosmos, The Pennsylvania State University, 525 Davey Lab, University Park, PA 16802, USA}

\author[0000-0001-9616-1789]{Hugh W. Sharp}
\affiliation{Department of Physics, 196A Auditorium Road, Unit 3046, University of Connecticut, Storrs, CT 06269, USA}

\author[0000-0001-8433-550X]{Matthew J. Temple}
\affil{Instituto de Estudios Astrof\'isicos, Facultad de Ingenier\'ia y Ciencias, Universidad Diego Portales, Av. Ej\'ercito Libertador 441, Santiago, Chile}

\author[0000-0002-6893-3742]{Qian Yang}
\affiliation{Department of Astronomy, University of Illinois at Urbana-Champaign, Urbana, IL 61801, USA}
\affiliation{Center for Astrophysics | Harvard \& Smithsonian, 60 Garden Street, Cambridge, MA 02138, USA}

\author[0000-0002-7817-0099]{Grisha Zeltyn}
\affiliation{School of Physics and Astronomy, Tel Aviv University, Tel Aviv 69978, Israel}

% Gave Comments (alphabetically)

\author[0000-0002-3601-133X]{Dmitry Bizyaev}
\affiliation{Apache Point Observatory and New Mexico State University, P.O. Box 59, Sunspot, NM, 88349-0059, USA}
\affiliation{Sternberg Astronomical Institute, Moscow State University, Moscow}

\begin{abstract}
We present a high-cadence multi-epoch analysis of dramatic variability of three broad emission lines (\MgII, \Hb, and \Ha) in the spectra of the luminous quasar (\LumCont\ = \targetLum) SDSS J141041.25+531849.0  at $z = 0.359$ with 127 spectroscopic epochs over 9 years of monitoring (2013-2022). 
% The variability of each line is quantified using a nonparametric description, although our conclusions remain consistent when fitting the emission lines with other methods.
We observe anti-correlations between the broad emission-line widths and flux in all three emission lines, indicating that all three broad emission lines ``breathe'' in response to stochastic continuum variations. We also observe dramatic radial velocity shifts in all three broad emission lines, ranging from $\Delta{v}$ $\sim$400~km~s$^{-1}$ to $\sim$800~km~s$^{-1}$, that vary over the course of the monitoring period. 
% We reject a supermassive black hole binary explanation for the radial velocity variation due to the implausibility of a putative binary period that produces the recurring velocity shifts without destroying the broad line region. 
% Instead, our 
Our preferred explanation for the broad-line variability is complex kinematics in the broad-line region gas. We suggest a model for the broad-line variability that includes a combination of gas inflow with a radial gradient, an azimuthal asymmetry (e.g., a hot spot), superimposed on the stochastic flux-driven changes to the optimal emission region (``line breathing''). Similar instances of line-profile variability due to complex gas kinematics around quasars are likely to represent an important source of false positives in radial velocity searches for binary black holes, which typically lack the kind of high-cadence data we analyze here. The long-duration, wide-field, and many-epoch spectroscopic monitoring of SDSS-V BHM-RM provides an excellent opportunity for identifying and characterizing broad emission-line variability, and the inferred nature of the inner gas environment, of luminous quasars.

\end{abstract} 
\section{Introduction}
Active Galactic Nuclei (AGN) are powered by supermassive black holes that are actively accreting matter at the centers of massive galaxies \citep{LyndenBell1969}. A hallmark signature of many AGN is the existence of broad emission lines, as first described by \cite{Seyfert1943}. Such lines arise from photoionization of the gas of the broad-line region (BLR), which is a distribution of gas that is thought to be moving in virialized orbits \citep[]{Bentz2009, Shapovalova2010, Barth2011a, Barth2011b, Grier2013a} close to the central black hole. As the radius of the BLR is on the order of light-days, it is difficult to spatially resolve with current technology. To date, there have been only a handful of studies that have been able to spatially resolve the BLR using near-infrared interferometry \citep{GravityCollab2020}. Investigating the structure of the BLR generally requires indirect techniques like reverberation mapping \citep[]{Blandford1982, Peterson1993,Cackett2021}.

Reverberation mapping (RM) is a technique that utilizes the fact that variations in the continuum flux of the AGN are followed by variations in the broad-emission lines from the BLR, with a time delay, $\tau$, that corresponds to the light-travel time between the continuum-emitting region and the BLR. The time delay is thus related to the typical radius of the BLR by the relation $R_{\rm BLR} = c \tau$. If we assume that the BLR is virialized, then the mass of the central black hole is determined by the virial product, namely,

\begin{equation}
    \label{BH_mass}
    M_{\rm{RM}} = f \frac{v^{2} R_{\rm BLR}}{G}
\end{equation}
where $v$ is the velocity of the BLR gas, $R_{\rm BLR}$  is the BLR radius, and $G$ is the gravitational constant. A dimensionless factor $f$ is introduced into the equation to parameterize the inclination angle,
% which is unknown, and the detailed geometry
unknown geometry, and orientation
of the BLR. The dimensionless factor $f$ is calibrated by comparison with spatially resolved BLR measurements \citep{Sturm2018}, the kinematics of stars and gas near the AGN \citep[e.g.,][]{Grier2013b, Woo2015}, and/or velocity-resolved reverberation mapping \citep[e.g.,][]{Pancoast2014a,Grier2017a}.

Both observations \citep[]{Wilhite2006, Park2012, Barth2015, Dexter2019, Homan2020, Wang2020} and photoionization modeling \citep[]{Korista2004, Cackett2006, Goad2014} have demonstrated that an increase in central flux from the AGN accretion disk results in an increase in the radius of the BLR ``optimal emitting region'' and a corresponding decrease in the orbital velocity of the gas since the line emission originates further out. This phenomenon is known as ``line breathing'' and manifests itself as an anti-correlation between the broad-line flux and width of the BLR over time. Quantifying the breathing behaviors of broad emission lines provides valuable constraints on the geometry, kinematics, and photoionzation of the BLR \citep{Wang2020}.

% The continuum flux of quasars exhibits variability of order $\Delta \rm{f} / \rm{f} \sim 10$\% \edit1{(f in this case refers to the continuum flux, not to be confused with the dimensionless factor f given in Equation~\ref{BH_mass})}

The continuum flux of quasars exhibits variability of order $\sim$10\% on timescales of weeks to years that is thought to be driven by thermal fluctuations in the accretion disk \citep[]{Ulrich1997, VandenBerk2004, Kelly2009, MacLeod2010}. Periodicity in quasar photometric lightcurves has been a popular method for searching for SMBH binary candidates \citep[]{Valtonen2008, Graham2015, Ackermann2015, Li2016, Charisi2016, Sandrinelli2018, Severgnini2018, Li2019, Chen2020, Liao2021, Zhang2022}. However, \cite{Vaughan2016} demonstrated that the stochastic variability of single (non-binary) quasars can resemble a periodic signal that can span a few periods, leading to false-positive identification of SMBH binaries. An alternative method for identifying binary SMBHs is searching for periodic broad emission-line radial-velocity shifts \citep[]{Gaskell1983, Loeb2010, Eracleous2012, Shen2013, Liu2014, Runnoe2017, Guo2019} akin to spectroscopic binary stars. This method requires high-cadence, long-term spectroscopic monitoring as the binary orbits are on the order of decades \citep{Yu2002}.

\begin{figure*}[t]%[ht!]
\epsscale{1.1}
\plotone{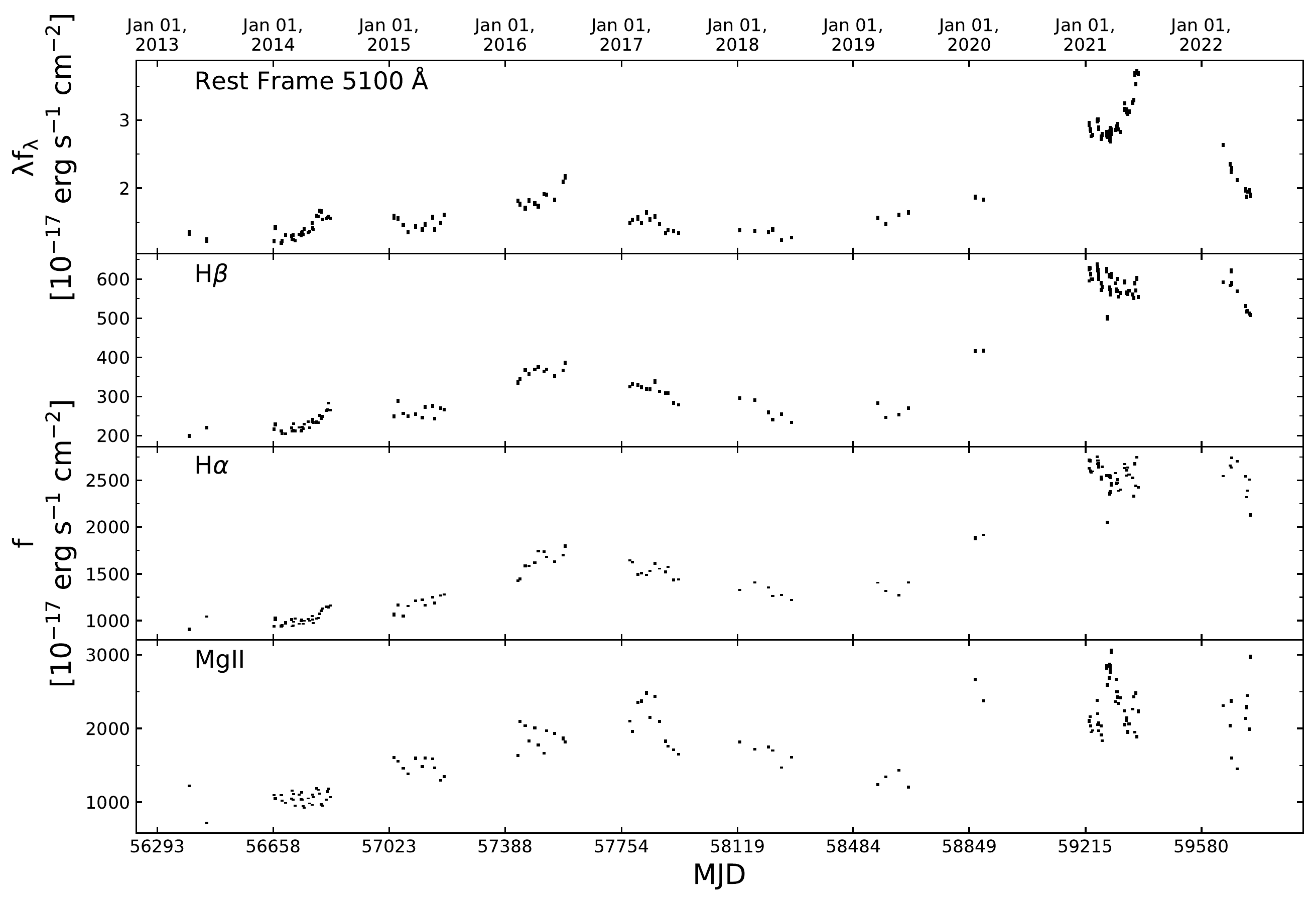}
\figcaption{Spectroscopic light curves for the rest-frame 5100\angstrom\ continuum and the three emission lines (\Hb, \Ha, \MgII). All 3 of the lines are brightest in the 2021-2022 monitoring period, although \Hb\ and \Ha\ have more extreme brightening than \MgII. The emission-line variability generally appears to follow the continuum variability with lags consistent with the previously measured $\tau \sim 30$~days for the Balmer lines \citep{Grier2017b} and $\tau \sim 145$~days for \MgII\ \citep[]{Homayouni2020}.
\label{Fig:light_curve}}
\end{figure*}

Radial velocity shifts of broad emission lines can also result from recoiling BHs \citep{Herrmann2007, Eracleous2012}, tidal disruption events \citep{Gezari2021}, and gas outflows/inflows \citep{Brotherton1994, Storchi-Bergmann2010, Rakshit2018, Kovacevic2022}. In addition, \cite{Barth2015} found that \Hb\ velocity centroids can undergo dramatic changes, on timescales of a month, in response to continuum flux variations. The radial velocity shifts in this case are a product of asymmetric reverberation by the BLR, and can appear as false-positive detections in binary black hole searches.

In this paper, we present observations of an AGN with unusual broad emission-line variability, SDSS~\targetLong\ (hereafter \target), found within a broader search for variability in broad emission-line profiles in the recently started Sloan Digital Sky Survey V (SDSS-V, \citealt{Kollmeier2017}, \citealt{Zasowski2023}). Section~\ref{Sec2} describes the sample selection, our criteria to identify quasars with unusual variability in their broad emission-line shapes, and provide the general characteristics of the object of interest. Section~\ref{Sec3} describes the 
% process with which we quantified 
methods we use to quantify
the broad emission-line profiles. Section~\ref{Sec4} describes the broad emission-line profile changes and presents a physical model to explain the observations. Section~\ref{Sec5} summarizes our results.

Throughout this work, we assume a \cosmo.

\section{Observations and Parent Sample}
\label{Sec2}
\subsection{Data}
\label{Data}
The data are from the third \citep{Eisenstein2011}, fourth \citep{Blanton2017}, and the ongoing fifth generation (\citealt{Kollmeier2017}, \citealt{Zasowski2023}) of the Sloan Digital Sky Survey (SDSS, \citealp{York2000}). The data were obtained using the plate-based, fiber-fed BOSS spectrograph \citep{Smee2013} mounted on the 2.5m SDSS telescope \citep{Gunn2006} at the Apache Point Observatory. The spectrograph has a dual-channel design, a blue channel (3000\angstrom\ < $\lambda$ < 6350\angstrom) and a red channel (5650\angstrom\ < $\lambda$ < 10,400\angstrom), both with a spectral resolution of $R \sim 2000$. The SDSS-III and SDSS IV data (2013-2020) were reduced with the v5\_13\_0 version of idlspec2d and the SDSS-V data (2021-present) were reduced with the v6\_0\_9 version of idlspec2d, the SDSS BOSS spectroscopic reduction pipeline \citep{Bolton2012}.

The spectroscopic monitoring spans a range of 9 years (2013 - 2022) with 127 epochs. An `epoch' generally represents observations taken in a single night, but in some cases epochs will include observations coadded from up to 3 consecutive nights in order to pass the `epoch-completion threshold' defined as a blue-channel based signal-to-noise ratio (SNR) threshold of ${\rm SNR}^2(g) > 20$ for a target of fiducial point spread function (PSF) magnitude $g=22$. This observing design aims to maintain a constant and useful SNR for all epochs, although some epochs have lower SNR because they could not be completed (i.e., pass the epoch completion threshold) within 3 nights. 

Figure~\ref{Fig:light_curve} shows the spectroscopic light curves \footnote{The observables shown here are explained further in Section~\ref{RM160_characteristics}} for our quasar of interest, \target, as a demonstration of the SDSS Reverberation Mapping (SDSS-RM) and SDSS-V Black Hole Mapper Reverberation Mapping (BHM-RM) data sets. The highest density of monitoring from SDSS-III and SDSS-IV occurred in 2014 (30 epochs) and at the start of SDSS-V in 2021 (39 epochs). These light curves include a second-order calibration in flux and wavelength using the \OIII$\lambda$5007 narrow emission line (see Section~\ref{calibrations} for details).

\subsection{Parent Sample}
\label{Sample}
The parent sample for our broader unusual variability search consists of all 320 quasars that have been monitored by both the SDSS-V BHM-RM program (for details: see Trump et al. in prep.) and SDSS-RM (for details: see \citealp{Shen2015a, Shen2019b}). These targets lie within the SDSS-RM field, which is a 7 deg$^{2}$ field that has been observed as a part of SDSS-RM in SDSS-III and SDSS-IV from 2013-2020, and then by BHM-RM in SDSS-V since 2021 (with monitoring scheduled to continue through at least 2026). The parent sample spans a redshift range of $0.1~<~z~<~4.34$ and is magnitude-limited by $i_{\rm PSF} < 21.7$. The median redshift of the parent sample is $z_{\rm med} = 1.52$ and the median \textit{i}-band magnitude of the parent sample is $\textit{i}_{\rm med} = 21.09$.

\subsection{Identifying Unusual Line Profile Changes in SDSS-RM / BHM-RM Quasars}
\label{VariabilitySearch}
Our object of interest, \target, was found during a broad search for quasars with variability in their broad emission-line profiles. To measure variability, we quantified changes to the broad emission-line profiles in each epoch using the reduced chi-squared \reducedChi, with $\nu$ the degrees of freedom, $O_{i}$ the observed flux density, $M_{i}$ the model flux density (described below), and $U_{i}$ the observational uncertainty summed across a wavelength range indexed by $i$.

\begin{figure}[t]%[ht!]
\epsscale{1.1}
\plotone{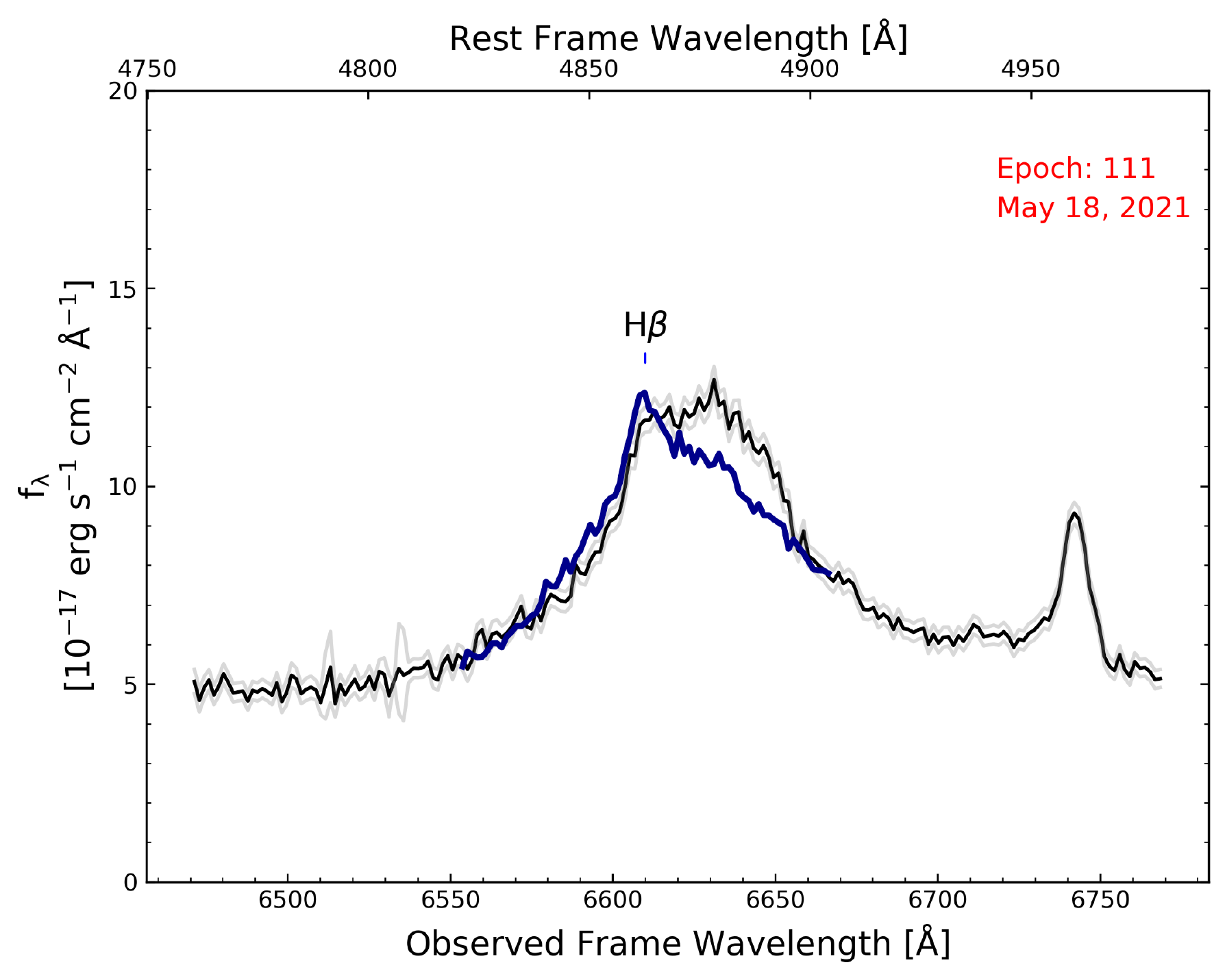}
\figcaption{The \Hb\ spectral region for \target\ in Epoch 111 (May 18, 2021 corresponding to an MJD of 59352). The dark blue line is the median spectrum computed for the \Hb\ line centered on the line center and has a total width of 1.5 $\times$ FWHM reported by \citet{Shen2019b}. In this particular epoch the \Hb\ emission-line profile is redder than the median profile.
% The online version of this Figure is an animation showing the \Hb\ region in all epochs.
The online version of this figure is an animation. The animation is 1:00 minutes long and shows the time evolution of \target\ for all 127 epochs of spectra beginning in April 11, 2013 and ending on June 03, 2022.
\label{Fig:GIFexample}}
\end{figure}

We use the median spectrum across all 127 epochs as our model flux density in our chi-squared calculation. We seek to identify changes in the line-profile \textit{shape}, rather than the (commonly observed) brightening or dimming of the overall broad line. To accomplish this, we allow the median spectrum in the line-profile region to scale up and down by a multiplicative factor computed from the ratio of the median flux density of the line-profile region at each epoch divided by the average flux density of the line-profile region across all epochs. Figure~\ref{Fig:GIFexample} shows an example spectrum for \target, where the blue line shows the median spectrum centered on the emission-line region and spanning a total width of 1.5 $\times$ FWHM reported in \cite{Shen2019b}, which is the extent to which we measure variability of the broad emission-line profile.

\begin{figure}[t]%[ht!]
\epsscale{1.1}
\plotone{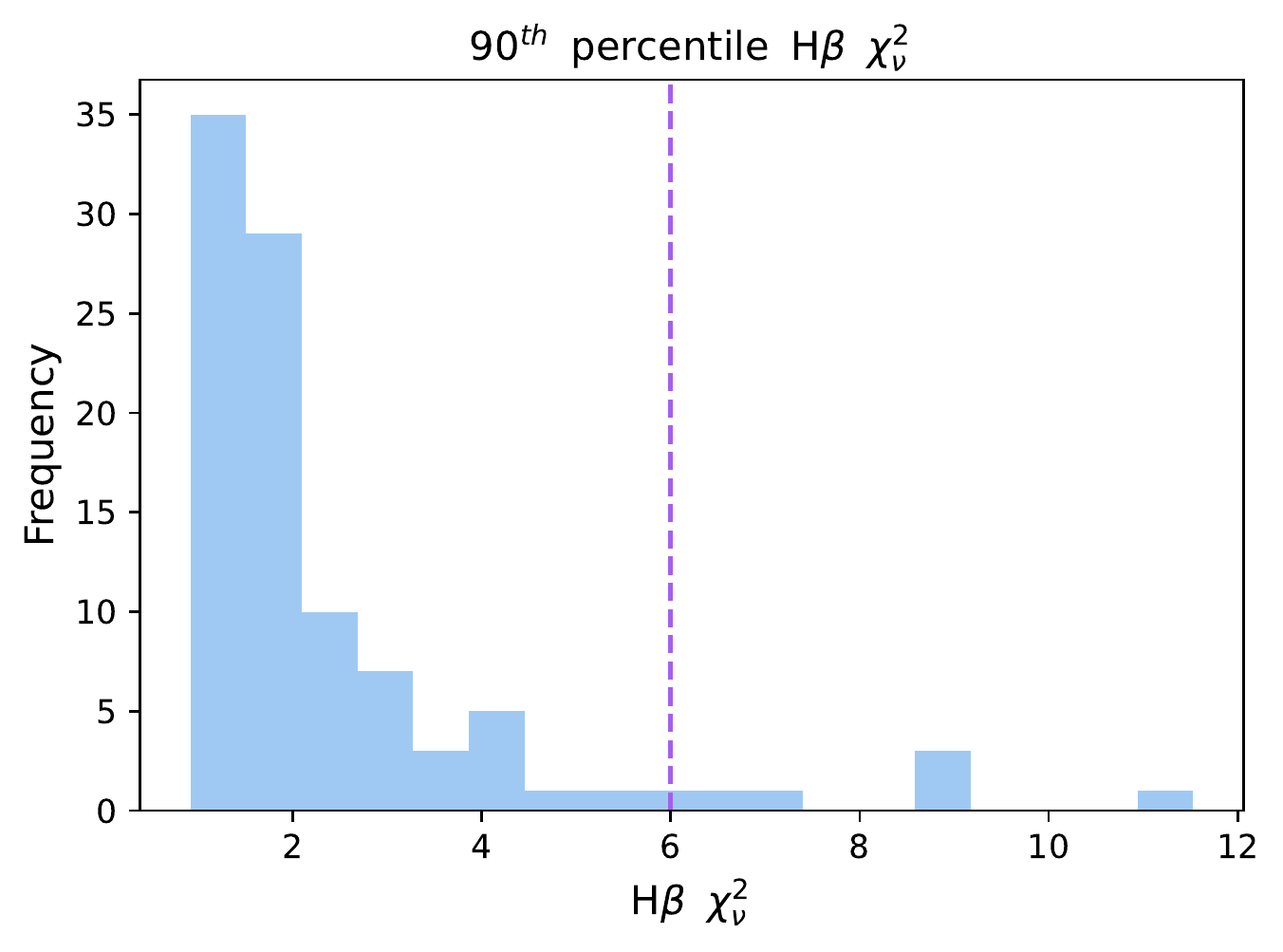}
\figcaption{Distribution of the \ninetychi\ measured from the multi-epoch \Hb\ region spectra for the 320 objects in our parent SDSS-RM sample. The vertical, dashed purple line indicates our criterion (\ninetychi\ $\geq$ 6) for identifying unusual line-profile variability. The \ninetychi\ for \target\ is 8.44 and was chosen as it had the highest SNR and the most apparent broad emission-line profile variability, while the other ones that satisfied our \ninetychi\ criterion were either noisy and/or had less apparent variability in their broad emission-line profiles.
\label{Fig:Hb90th}}
\end{figure}

We identified interesting candidates using the 90th percentile of the distribution of the $\chi_{\nu}^{2}$ values from each epoch, choosing \ninetychi\ > 6 as a threshold associated with the tail of line-profile variability. Figure~\ref{Fig:Hb90th} shows the distribution of \Hb\ \ninetychi\ of our sample.

There are 15 objects that fit our \ninetychi\ criterion. We visually inspected the variability of the 15 objects by creating animations of the time-variable spectra for all 127 epochs. Out of those 15 objects, we identified the subject of this study, \target. The \ninetychi\ for \target\ is 8.44. We noticed from visual inspection that the \Hb, \Ha, and \MgII\ broad emission lines of \target\ all appear to undergo significant velocity shifts over time. The apparent variability of the remaining objects were largely the result of noise.

\subsection{The Source of Interest: \target}
\label{RM160_characteristics}
The subject of this study is \target, a luminous quasar (\LumCont = \targetLum) in the SDSS-RM field. It has a redshift of $z = 0.359$ and an \textit{i}-band magnitude of $i = 19.68$. 

There are published \Hb, \Ha, and \MgII\ RM time lags for this object. The \Hb\ and \Ha\ lags were measured by \cite{Grier2017b} using only the 2014 data. The observed-frame \Hb\ lag is \GrierHbLagObs\ and the observed-frame \Ha\ lag is \GrierHaLagObs. The observed-frame \MgII\ lags were measured in \cite{Homayouni2020} using 4 years of data (2014-2017). The observed-frame \MgII\ lag is \HomayouniMgIILagObs. 
% \edit1{We note that the \MgII\ lag for this object is not in the "gold-sample" in \cite{Homayouni2020} which is defined as objects whose \MgII\ lags have false-positive rates of $\leq$10\%. \target's \MgII\ lag has a false-positive rate of 16\%. As such, the \MgII\ lag for \target\ may be unreliable and it is worth noting that none of our conclusions rely on an \textit{exact} value of the \MgII\ lag for this object.}
We note that the \MgII\ lag for this object has a false-positive rate of 16\%, which is not in the "gold-sample" (false-positive rate of $\leq$10\%) of \cite{Homayouni2020}. As such, the \MgII\ lag for \target\ may be unreliable. In general, we assume that the \MgII\ lag is longer than the lags of the Balmer lines (i.e., we assume the BLR is stratified; see \citealp{Clavel1991, Reichert1994}).
The black hole mass of \target\ was computed in \cite{Grier2017b} to be \targetBHMass.

From HST imaging (taken on 2020 September 28), the host-galaxy contribution for \target\ is 14\% in F606W \citep{Li2023}. This measurement does not take into account the 2\arcsec\ SDSS fiber which would make the host-fraction even smaller, thus we do not account for the marginal host contribution in our analysis and we assume that the spectrum is dominated by the quasar.

Figure~\ref{Fig:light_curve} shows the continuum and emission-line light curves for \target. The light curve behavior appears to be qualitatively consistent with the previously measured lags, with the Balmer-line variability appearing to follow the same pattern as the continuum after a lag of $\sim$30~days and the \MgII\ variability following the same pattern after an additional $\sim$145~days. 
%% YS: This may not be visually obvious to the reader or the referee.

\begin{figure*}[bt]%[ht!]
\epsscale{1.1}
\plotone{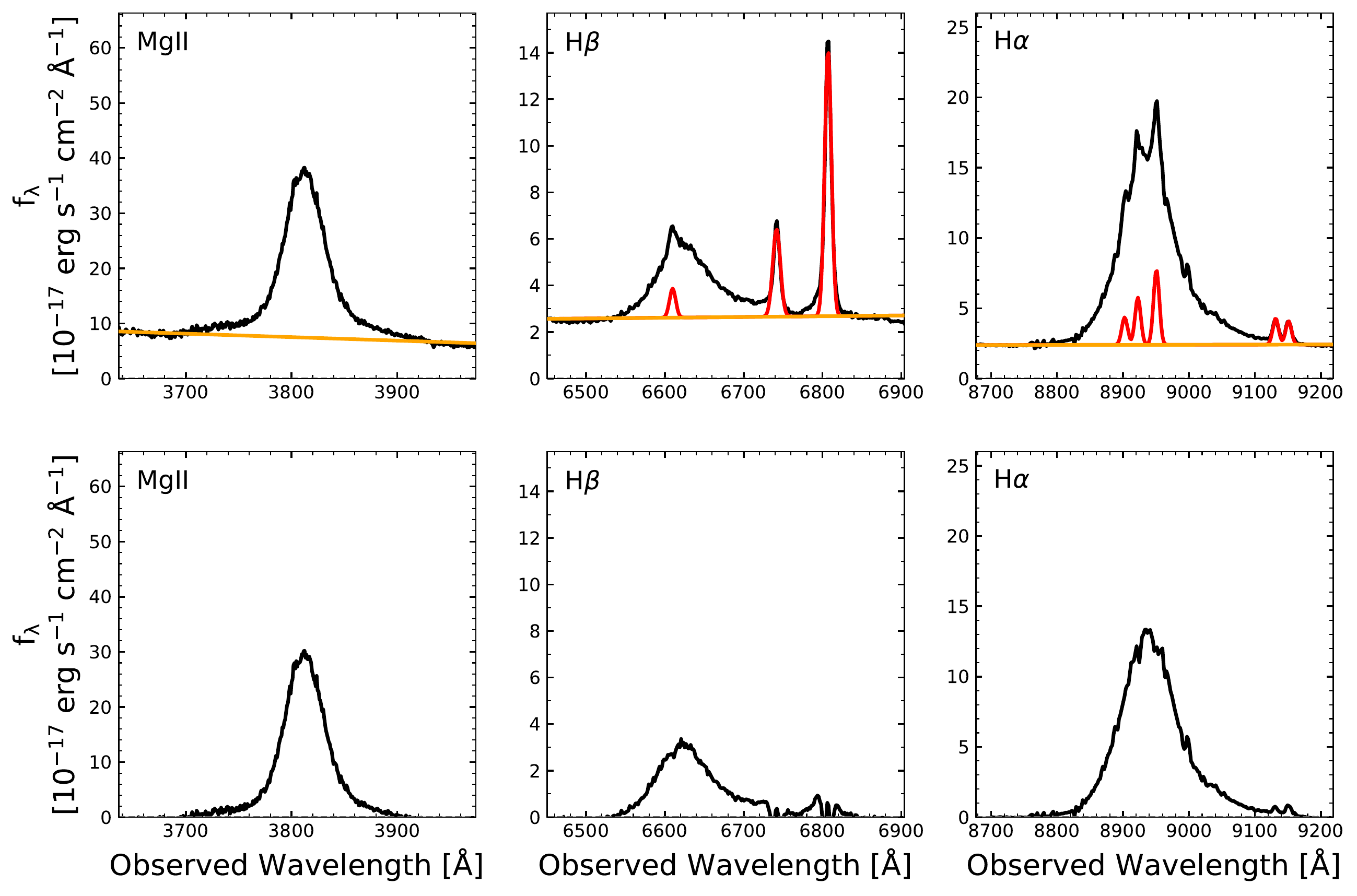}
\figcaption{Median spectra for \MgII, \Hb, and \Ha\ of \target. The top row shows the full spectrum in black along with the narrow emission-line fits in red and the continuum fits in orange. The bottom row shows the continuum and narrow-line subtracted median spectra for \MgII, \Hb, and \Ha, respectively. Our continuum and narrow-line subtracted spectra show clear broad-line profiles (the residuals from non-Gaussian narrow lines do not affect the broad-line measurements).
\label{Fig:CompositeSpectra}}
\end{figure*}

\section{Quantifying the Emission-Line Profiles}
\label{Sec3}
\subsection{Fitting the Continuum}
\label{continuum_fits}
To isolate and model the broad emission-line regions, we first need to subtract the continuum from the spectra. To subtract the continuum, we fit a first-order polynomial to the spectrum based on the median of the continuum over 50 pixels from the line-free regions blueward and redward of the broad lines (and nearby narrow lines for \Hb\ and \Ha). The continuum fits for the median spectrum are shown in yellow in the top panels in Figure~\ref{Fig:CompositeSpectra}. Appendix~\ref{AltFitting} presents alternative modeling with \texttt{PyQSOFit} \citep[]{Guo2018, Shen2019b} that separately fits the continuum and iron psuedo-continuum and finds consistent variability patterns in the fitted line properties.

\subsection{Fitting the Narrow Emission Lines}
\label{narrow_fits}
% We fit each narrow emission line in the \Ha\ and \Hb\ emission-line regions with a single Gaussian. Since $\OIII\lambda$5007 does not vary over timescales of a few years \citep[]{Foltz1981, Peterson1982}, we used it as a benchmark to inform the fits of the other narrow lines.

We fit each narrow emission line in the \Ha\ and \Hb\ regions with a single Gaussian. We assume that the narrow lines are constant over the course of our monitoring period of $\sim$9 years. In Section~\ref{calibrations}, we confirm this assumption by demonstrating that the $\OIII\lambda$5007 narrow-line flux is constant over the monitoring period.

% We assume that the $\OII\lambda$5007 line is constant over our monitoring period of $\sim$9 years.

% We note that in the case of NGC 5548, the $\OIII\lambda$5007 line has been shown to vary over a timescale of $\sim$30 years \citep{Peterson2013}.}

% \edit1{We fit each narrow emission line in the \Ha\ and \Hb\ regions with a single Gaussian. We assume the $\OIII\lambda$5007 line is constant over our monitoring period of $\sim$9 years. We note that in the case of NGC 5548 the $\OIII\lambda$5007 line has been shown to vary over a timescale of $\sim$30 years \citep{Peterson2013}. In Figure~\ref{Fig:OIIIParams}, we show our single Gaussian fit parameters (flux, line center, and line width) for $\OIII\lambda$5007 over the course of our monitoring period and we confirm that $\OIII\lambda$5007 is not variable during our monitoring period.}

We used the $\OIII\lambda$5007 parameters to constrain the other narrow-line fits.
% We fit $\OIII\lambda$5007 using a single Gaussian and used the parameters (line width and line center) to constrain the other narrow-line fits. 
Specifically, we tied the line centers of the narrow-line Gaussian fits, for each epoch, to $\OIII\lambda$5007 using the narrow-line wavelength centers from \cite{VandenBerk2001}. We also tied the line widths of the narrow-line Gaussian fits, for each epoch, to the line width from the $\OIII\lambda$5007 fit. The $\OIII\lambda$4959 line flux was also constrained to be 1/3 that of the $\OIII\lambda$5007 line flux \citep{Storey2000}.

% \begin{figure*}[t!]%[ht!]
% \epsscale{1.1}
% \plotone{medianSpectrumComposite_fixed.pdf}
% \figcaption{Median spectra for \MgII, \Hb, and \Ha of \target. The top row shows the initial spectrum in black for \MgII, \Hb, and \Ha\, respectively, along with the narrow emission-line fits in red and the continuum fits in orange. The bottom row shows the continuum and narrow-line subtracted median spectra for \MgII, \Hb, and \Ha, respectively. Our continuum and narrow-line subtracted method results in clear broad-line profiles (the residuals from non-Gaussian narrow lines do not affect the broad-line measurements).
% \label{Fig:CompositeSpectra}}
% \end{figure*}

For each individual fit for the \Ha\ region, we fit the \Ha\ narrow line without constraints on the amplitude and found that this resulted in a clean-looking residual (i.e., a clear broad emission line profile with the narrow emission lines cleanly subtracted out). However, this unconstrained approach to the narrow \Hb\ line resulted in poor fits with large apparent residuals. Therefore, we constrained the amplitude of the narrow \Hb\ line to a value that produced a smooth broad-line residual that lacked a cuspy narrow-line peak in the median spectrum fit. We then applied that narrow \Hb\ amplitude to the fits for all epochs. 

\begin{figure*}[t]%[ht!]
\epsscale{1.1}
\plotone{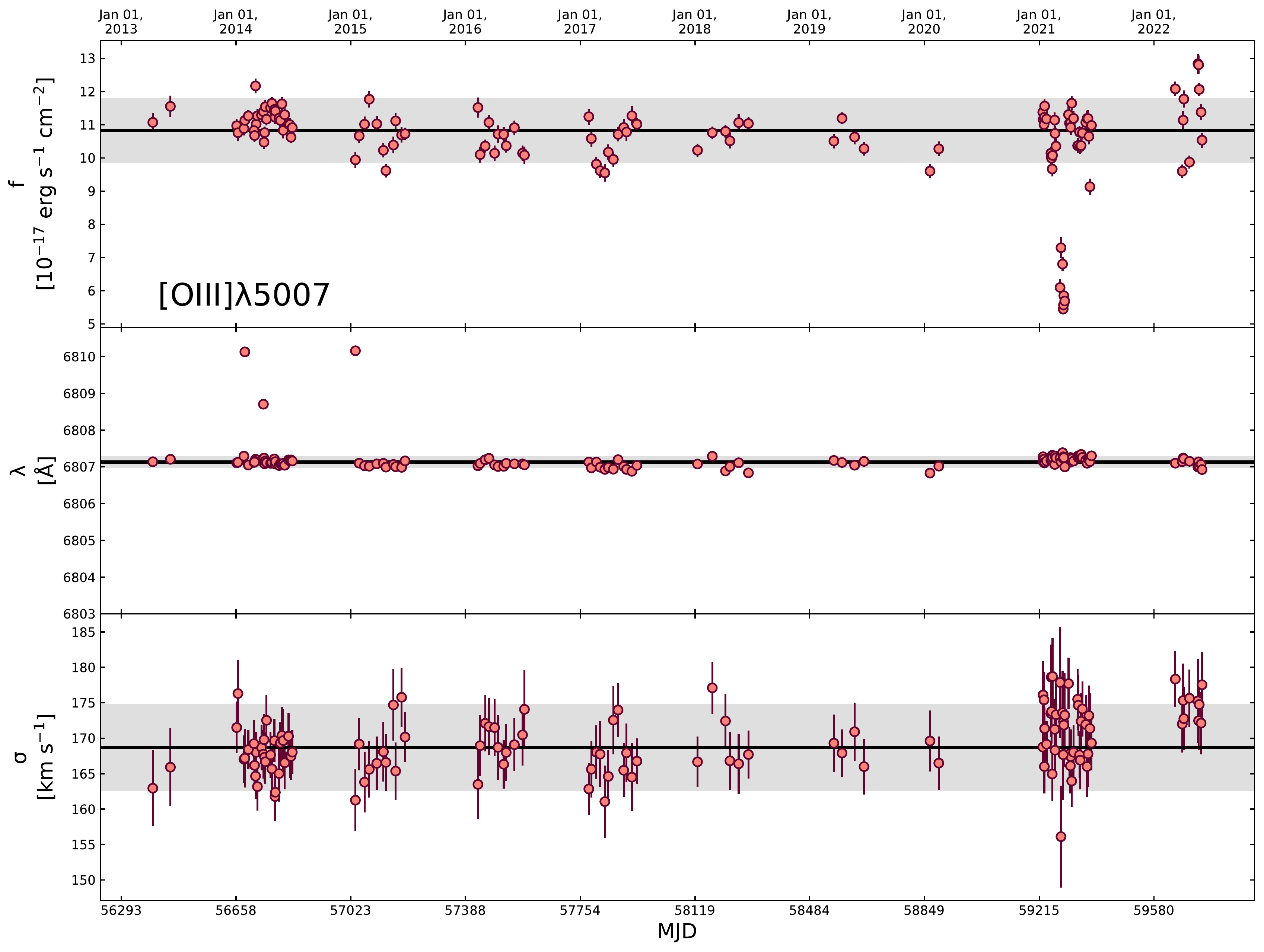}
\figcaption{The \OIII$\lambda$5007 flux (top), line center (middle), and line width $\sigma$ (bottom) measured from single-Gaussian fits of the continuum subtracted spectra at each epoch. In all panels, the black line indicates the median and the gray shading represents the NMAD. We use the apparent changes in $\OIII\lambda$5007 flux and line center for a second-order flux and wavelength calibration for the spectra of each epoch.
\label{Fig:OIIIParams}}
\end{figure*}

Figure~\ref{Fig:CompositeSpectra} illustrates the process of fitting and subtracting both the continuum and the narrow emission lines. There are limitations to modeling complex spectra with a single Gaussian. For example, the $\OIII\lambda$5007 line appears to have a blue wing that cannot be modeled using a single Gaussian and would be better fit with two or more Gaussians. This asymmetric, non-Gaussian profile has been shown to be due to outflowing, ionized gas in the narrow-line region (NLR) \citep[]{Rojas2020, Ayubinia2022, Molina2022}.  However, these small residuals caused by non-Gaussian shapes in the narrow emission lines do not affect the measured broad-line profiles. We visually inspected the fits for each broad emission line in each epoch and confirmed that our method produces clean broad-line profiles that have effective subtractions of the continuum and narrow emission lines.

\subsection{Second-Order Calibrations Based on Narrow Emission Lines}
\label{calibrations}
We investigated the stability of the narrow emission lines to examine and improve the flux and wavelength calibration. The $\OIII\lambda$5007 line is observed to be stable over timescales of a few years \citep[]{Foltz1981, Peterson1982} and so it is often used as a flux-calibration standard in AGN spectra. We examine the possibility of applying a spectrophotometric calibration by fitting the $\OIII\lambda$5007 line with a single Gaussian in each epoch, which is separate from the aforementioned fitting procedure in Section~{\ref{narrow_fits}}. In Figure~\ref{Fig:OIIIParams}, we show the Gaussian fit parameters for $\OIII\lambda$5007 line flux, center, and width, as well as the respective median values (black) and normalized median absolute deviations (NMAD; gray regions). Figure~\ref{Fig:OIIIParams} confirms our assumption of a non-variable $\OIII\lambda$5007 line throughout the course of our monitoring period. We note that in the case of NGC 5548, the $\OIII\lambda$5007 line has been shown to vary over a timescale of $\sim$30 years by $\sim$10\% \citep{Peterson2013}.

% \begin{figure*}[t]%[ht!]
% \epsscale{1.1}
% \plotone{OIIIParamsVsMJD.pdf}
% \figcaption{The \OIII$\lambda$5007 flux (top), line center (middle), and line width $\sigma$ (bottom) measured from single-Gaussian fits of the continuum subtracted spectra at each epoch. In all panels, the black line indicates the median and the gray shading represents the NMAD. We use the apparent changes in $\OIII\lambda$5007 flux and line center for a second-order flux and wavelength calibration for the spectra of each epoch.
% \label{Fig:OIIIParams}}
% \end{figure*}

Since the $\OIII\lambda$5007 narrow line does not vary over these timescales, the changes in $\OIII\lambda$5007 flux represent epoch-dependent changes in the spectrophotometric calibration, with 8 epochs in 2021 that fall well below the median. The $\OIII\lambda$5007 line center shifts by exactly 1 or 2 pixels in 3 epochs, indicating a shift in the wavelength calibration. The $\OIII\lambda$5007 line width is constant within its uncertainties, indicating that the spectral resolution is stable throughout the observations. 

We observe similar changes in the fitted line fluxes and centers for the \OII\ and \SII\ emission lines (with larger uncertainties for these weaker lines). This suggests that the flux and wavelength changes are gray (not wavelength-dependent) and systematic (not intrinsic to \target). In other words, the flux and wavelength changes observed for $\OIII\lambda$5007 represent gray calibration issues for the entire spectrum.

We perform a second-order flux and wavelength calibration that forces the $\OIII\lambda$5007 flux and wavelength to be constant across all epochs and apply it to the spectra at each epoch. We scale the spectrum at each epoch by a factor of $\tilde{f}\OIII / f\OIII$ and we correct the wavelength of each spectrum by a factor of $\tilde{\mu}\OIII / \mu\OIII$ , where $\tilde{f}\OIII$ and $\tilde{\mu}\OIII$ are the median $\OIII\lambda$5007 flux and line center, respectively across all epochs and $f\OIII$ and $\mu\OIII$ are the $\OIII\lambda$5007 flux and line center at each epoch.

\subsection{Quantifying the Broad Emission-Line Profile Variability}
\label{non_parametric}
We measure the broad emission-line properties of the continuum and narrow-line subtracted spectra using the moments of a distribution:

\begin{itemize}
    \item Line Flux: $f = \sum_{\lambda_{1}}^{\lambda_{2}} {f_{\lambda}(\lambda) \Delta \lambda} $
    \item Line Center: $\lambda_{\rm{C}} = \frac{\sum_{\lambda_{1}}^{\lambda_{2}} {\lambda f_{\lambda}(\lambda) \Delta \lambda}}{\sum_{\lambda_{1}}^{\lambda_{2}} {f_{\lambda}(\lambda) \Delta \lambda}}$
    \item Line Width: $\rm{\sigma} = \sqrt{\frac{\sum_{\lambda_{1}}^{\lambda_{2}} {(\lambda_{\rm{C}} - \mu)^{2} {\textit{f}_{\lambda}(\lambda) \Delta \lambda}}}{\sum_{\lambda_{1}}^{\lambda_{2}} {\textit{f}_{\lambda}(\lambda) \Delta \lambda}}}$
    \item FWHM: We measured FWHM by first applying a median boxcar smoothing of 5 pixels to our spectra at each epoch. We then used the FWHM routine from the \texttt{specutils} package \citep{Earl2022} to compute the FWHM.
\end{itemize}
Here $f_{\lambda}$ is the flux density, $\lambda$ is the wavelength, $\Delta \lambda$ is the wavelength per pixel, and $\lambda_1$ and $\lambda_2$ for each line are from \citet{VandenBerk2001}. These non-parametric measurements were chosen as they make no assumptions about the underlying shape of the broad line profiles. We used the non-parametric model on the continuum and narrow-line subtracted spectra (bottom row panel in Figure \ref{Fig:CompositeSpectra}).

%%%%% UNCOMMENT IF GOING BACK TO NAMED EQUATIONS.

% where $F_{\lambda}$ is the flux density, $\lambda$ is the wavelength and $\Delta \lambda$ is the spacing in the wavelength array. We also measure the FWHM of the emission line by applying a median boxcar smoothing of 5 pixels to our spectra. We then used the FWHM routine from the \texttt{specutils} package \citep{Earl2022} to compute the FWHM. These non-parametric measurements were chosen as they make no assumptions about the underlying shape of the broad line. We used the non-parametric model on the continuum and narrow line subtracted spectra (bottom row panel in Fig. \ref{Fig:CompositeSpectra}).

We employ a Monte Carlo resampling approach to estimate the uncertainties in line flux, line-center, and line width (both $\sigma$ and FWHM). For each epoch, we measured these quantities for 200 resampled spectra using the corresponding error spectrum. We adopted the standard deviations in the measured parameters from the 200 re-sampled spectra as our uncertainties for each epoch. 

\begin{figure*}[t]%[ht!]
\epsscale{1.1}
\plotone{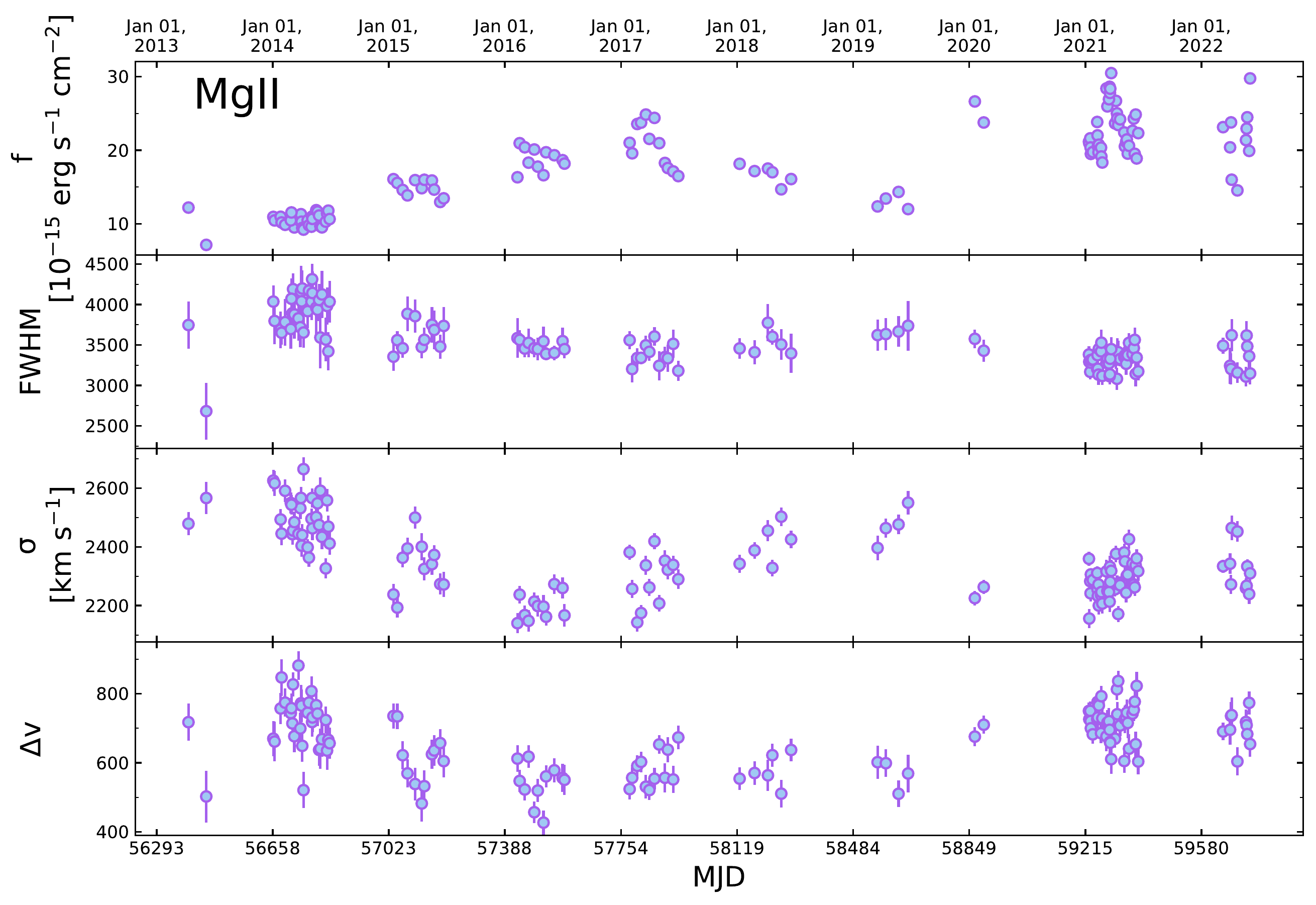}
\figcaption{Variability of the \MgII\ broad emission-line profile, as quantified by our non-parametric measurements of the line flux (top row), FWHM (2nd row), line-width $\sigma$ (3rd row), and the line-center velocity shift $\Delta{v}$ (bottom row). The units for FWHM, line-width $\sigma$, and line-center velocity shift $\Delta{v}$ are all in km~s$^{-1}$. The \MgII\ line gets brighter and narrower, fainter and broader and then brighter and narrower again. Meanwhile, the line-center starts very red (compared to the systemic redshift from the narrow emission lines), gets bluer and then gets as red as it was at the beginning. The velocity shifts in line-center do not appear to be simultaneously correlated with the changes in flux and line-width.
\label{Fig:nonParaMgIILineProfile}}
\end{figure*}

\begin{figure*}[t]%[ht!]
\epsscale{1.1}
\plotone{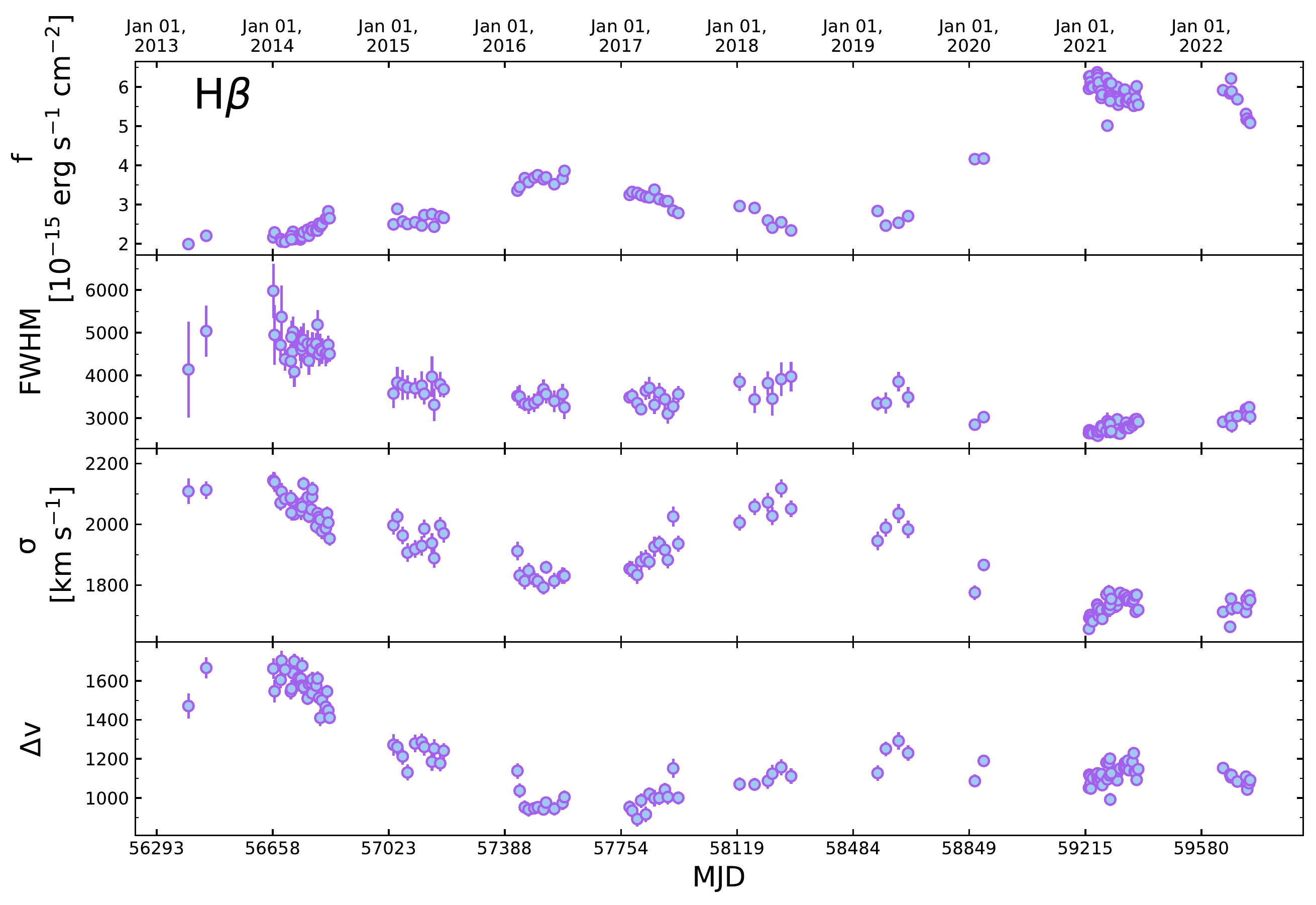}
\figcaption{Variability of the \Hb\ broad emission-line profile, as quantified by our non-parametric measurements of the line flux (top row), FWHM (2nd row), line-width $\sigma$ (3rd row), and the line-center velocity shifts $\Delta{v}$ (bottom row). The units for FWHM, line-width $\sigma$, and line-center velocity shift $\Delta{v}$ are all in km~s$^{-1}$. The \Hb\ line gets brighter and narrower, fainter and broader and then brighter and narrower again. Meanwhile, the line-center starts very red (compared to the systemic redshift from the narrow emission lines), gets bluer and then gets more redder and seems to plateau after the 2019 monitoring period. The velocity shifts in line-center do not appear to be simultaneously correlated with the changes in flux and line-width.
\label{Fig:nonParaHbLineProfile}}
\end{figure*}

\begin{figure*}[t]%[ht!]
\epsscale{1.1}
\plotone{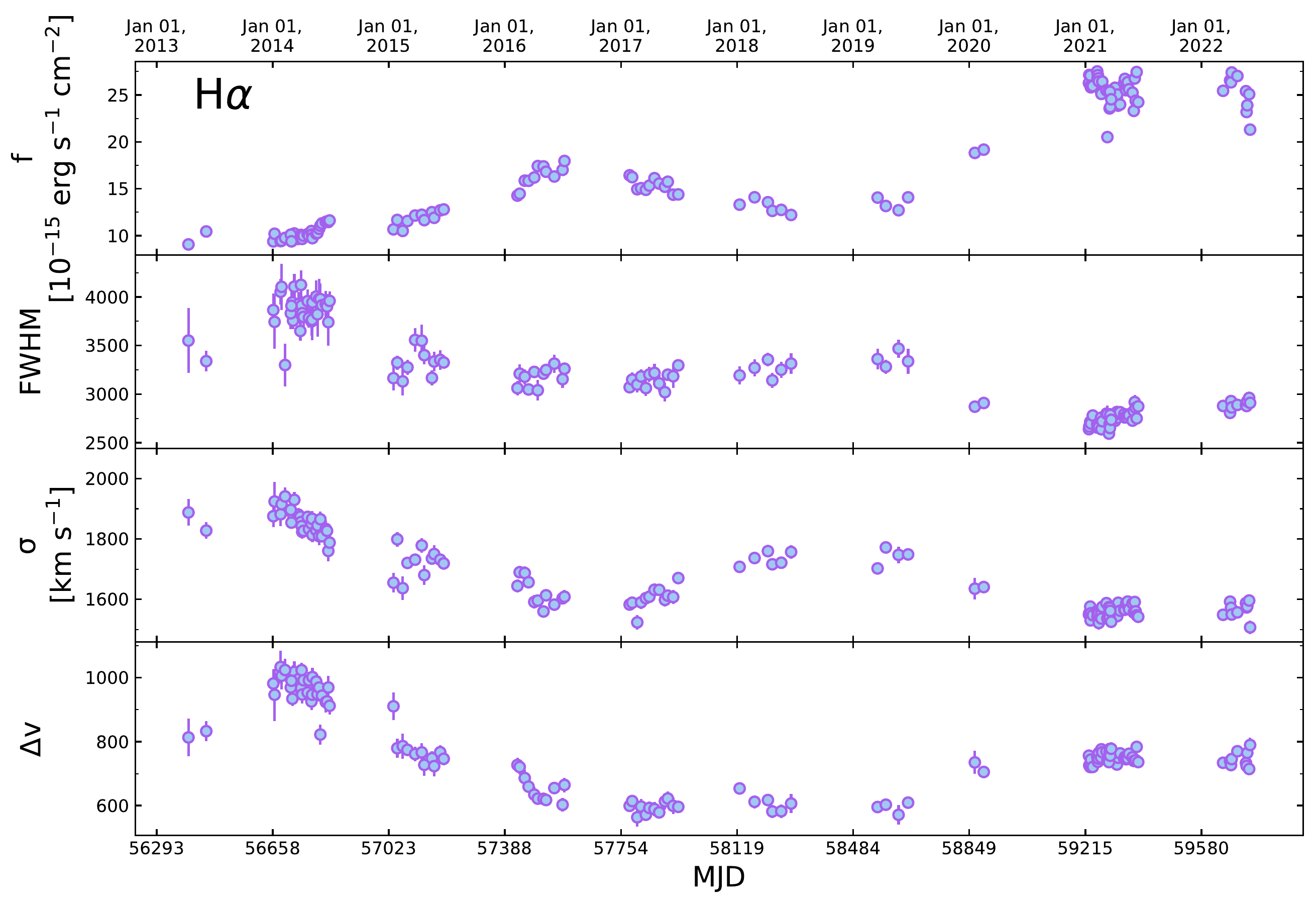}
\figcaption{Variability of the \Ha\ broad emission-line profile, as quantified by our non-parametric measurements of the line flux (top row), FWHM (2nd row), line-width $\sigma$ (3rd row), and the line-center velocity shifts $\Delta{v}$ (bottom row). The units for FWHM, line-width $\sigma$, and line-center velocity shift $\Delta{v}$ are all in km~s$^{-1}$. The \Ha\ line gets brighter and narrower, fainter and broader and then brighter and narrower again. Meanwhile, the line-center starts very red (compared to the systemic redshift from the narrow emission lines), gets bluer and then gets more redder, but not as red as it was in the 2014 monitoring period. The velocity shifts in line-center do not appear to be simultaneously correlated with the changes in flux and line-width.
\label{Fig:nonParaHaLineProfile}}
\end{figure*}

To test the robustness of our non-parametric measurements, we employed two alternative parametric fitting procedures: (1) using single Gaussians to model the broad emission lines and (2) using \texttt{PyQSOFit} which models the Fe\,\textsc{ii} psuedo-continuum and fits multiple Gaussians to the broad and narrow emission lines. The different methods all show the same relative variability of the broad-line profiles, with the exception of the $\sigma$ measurements from \texttt{PyQSOFit}: for details, see Appendix~\ref{AltFitting}. For clarity, we use the nonparametric measurements for our analysis throughout the text.

\section{Results and Discussion}
\label{Sec4}
\subsection{Broad Emission-Line Profile Variability}
\label{ProfileVariability}

Figures~\ref{Fig:nonParaMgIILineProfile}, \ref{Fig:nonParaHbLineProfile}, and \ref{Fig:nonParaHaLineProfile} show the variability of the \MgII, \Hb, and \Ha\ broad emission-line profiles as quantified by our non-parametric measures of flux, line width (FWHM and $\sigma$), and line center. In general, in each of the figures, the broad emission lines get brighter and narrower, fainter and broader, and then brighter and narrower again, with an anti-correlation between the flux and line width. This phenomenon is known as `line breathing' and we discuss it in more detail in Section~\ref{LineBreathing}. We note that the flux and line-width (FWHM and $\sigma$) of \MgII\ appear to vary as a lagged version of the Balmer lines, consistent with the observed lags in each emission line of this quasar presented in \cite{Grier2017b} and \cite{Homayouni2020}. This suggests that the MgII-emitting region is further away from the central engine, compared with the Balmer-emitting regions.

All three broad emission lines have line centers that are much redder than the systemic redshift, as determined from the narrow emission lines. The broad-line centers all have a similar general variability pattern of starting red, shifting bluer over a few years, and then getting redder near the end of our monitoring period. The broad \Hb\ line is redder than the other broad lines and has the largest radial velocity shifts (maximum change in $\Delta{v}$ of $\sim$800~km~s$^{-1}$, compared to $\sim$400~km~s$^{-1}$ for \Ha\ and \MgII). All three lines become bluest (but still redder than the systemic narrow-line redshift) in $\sim$2017 and become red again by $\sim$2020, with \MgII\ returning to its initial red center while \Ha\ and \Hb\ do not become as red as when they started.

Unlike the light curves of the emission-line flux, the comparison between the radial velocity shifts in the Balmer lines (\Hb\ and \Ha) and \MgII\ is inconsistent with the measured lags reported by \cite{Grier2017b} and \cite{Homayouni2020}. Specifically, the \MgII\ radial velocity shifts do not appear to mirror the Balmer-line shifts after a lag of $\sim$100~days. Instead, the \MgII\ radial velocity shifts appear to be a smoother version of the Balmer-line shifts with no apparent lag between them.

The radial velocity shifts of \target\ are qualitatively similar to, but more extreme than, what is observed for the lower luminosity Seyfert 1 AGN discussed by \cite{Barth2015}. For example, the largest shifts reported by \cite{Barth2015} are for NGC~4593 with radial velocity shifts of 266~$\pm$~11~km~s$^{-1}$ for the broad \Hb\ emission line. \target\ has a much more dramatic \Hb\ velocity shift of $\sim$800~km~s$^{-1}$ that occurs over $\sim$4~years. \cite{Sergeev2007} studied NGC 5548 and found radial velocity shifts of $\sim$1,000~km~s$^{-1}$ over a 30 year period.

% However, it has been shown that variability and luminosity are anti-correlated \citep{MacLeod2012}. So less luminous quasars like NGC 5548 are on average more variable than quasars like \target.

\subsection{Line Breathing}
\label{LineBreathing}
Line breathing is an anti-correlation between the flux and the width of a broad emission line over time that has long been predicted by photoionization modeling \citep{Korista2004}. Gas in a Keplerian orbit, as is the likely case for the BLR, has higher velocity at small radii and lower velocity at large radii. For BLR gas ionized by a central continuum, an increase in the continuum flux will over-ionize the gas nearest to the black hole and will thus increase the emissivity-weighted BLR radius. The increase in the optimal emitting radius will result in a decrease in the line width, since gas orbiting at a larger radius has a lower orbital velocity. Observations have shown that BLR line breathing can occur on timescales of days to weeks \citep{Barth2015}. Line breathing is described by the following relation:

\begin{equation}
    \label{lineBreathingEq}
    \Delta \, \log \, W = \beta \, \Delta \, \log \, L
\end{equation}
where $W$ is the line-width, $L$ is the luminosity and $\beta$ is some constant of proportionality. The broad-line luminosity is generally used for $L$ in Equation \ref{lineBreathingEq} as a (lagged) representation of the continuum luminosity that is responsible for driving the change in optimal emitting radius. Assuming a typical radius-luminosity relationship of $R_{\rm BLR} \propto L^{0.5}$ \citep{Bentz2013} and virial orbits ($v \propto R_{\rm BLR}^{-0.5}$), the expected relation between the changes in line-width and luminosity is $\Delta \, \textrm{log} \, W = \textrm{-0.25} \, \Delta \, \textrm{log} \, L$.

Reverberation mapping studies over time have shown that line breathing is observed for \Hb\ \citep{Cackett2006, Park2012, Barth2015, Wang2020}, typically consistent with the expected constant of proportionality of $\beta = -0.25$.
On the other hand, observations of \MgII\ have found weak or no anti-correlations between line width and flux \citep{Dexter2019, Homan2020, Yang2020, Wang2020}. In \CIV, some observations have found a ``reverse breathing'' effect whereby the line width of \CIV\ increases with increasing flux \citep{Wilhite2006, Wang2020}. Furthermore, \cite{Wang2020} found that \Ha\ shows much less breathing than \Hb\ on average. The differences in line breathing patterns between broad emission lines indicate that there could be a difference in the distribution of the gas around a particular ``optimal emitting region'' and also can indicate that the structure of the BLR is not uniform at all radii. Studying line breathing behavior for different emission lines can give clues about the multi-scale structure of the BLR.

\begin{figure*}[ht]
\epsscale{1.1}
\plotone{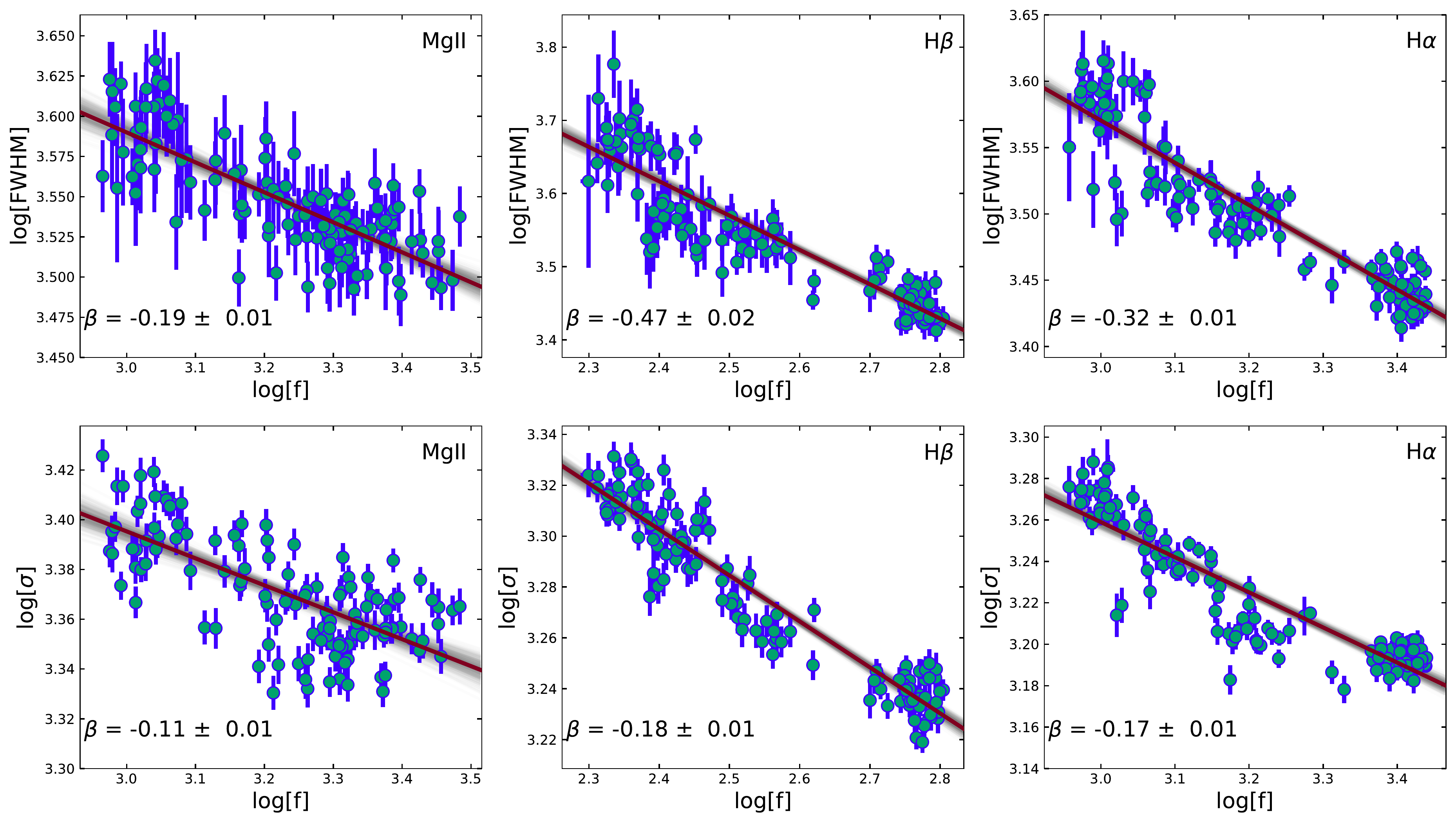}
\figcaption{Relationships between line width and line flux for \target. The green points are the data over our 9 year monitoring period. The maroon line is the best-fit linear relation and the gray shading indicates the distribution of best-fit lines from the Markov Chains used for the fitting. The top panels show the line FWHM vs.\ flux for \MgII, \Hb, and \Ha\ (left to right). The bottom panels show the line width computed from the second moment vs.\ flux for \MgII, \Hb, and \Ha\ (left to right). Line widths (FWHM and $\sigma$) are in units of km~s$^{-1}$ and line fluxes are in units of 10$^{-17}$~erg~s$^{-1}$~cm$^{-2}$.
\label{Fig:NonParaLineBreathing}}
\end{figure*}

We analyze line breathing for the three major broad lines accessible in \target: \MgII, \Hb, and \Ha\ over the entire 9 year monitoring period studied here (2013 - 2022). Figure \ref{Fig:NonParaLineBreathing} shows the relationship between broad-line width (FWHM and $\sigma$) and broad-line flux for \MgII, \Hb, and \Ha. We fit the lines with linear models, motivated by Equation~\ref{lineBreathingEq}, using the Bayesian linear regression package \texttt{linmix} \citep{Kelly2007}. The $\beta$ slopes determined from \texttt{linmix} are shown in Fig. \ref{Fig:NonParaLineBreathing} on the bottom left. All three emission lines exhibit an anti-correlation between broad-line width and broad-line flux, with a slope that is steeper for FWHM and shallower for $\sigma$. The \Hb\ line has the steepest anti-correlation and \MgII\ has the shallowest anti-correlation, in general agreement with previous work \citep{Wang2020}.

The line breathing properties of \target\ were previously measured by \cite{Wang2020} using the continuum flux and broad-line widths from 29 epochs of spectra observed over 2014-2017 (choosing only epochs that are 2$\sigma$ above the mean S/N of each season from the 64 total epochs in 2014-2017). Our study measures the line breathing properties over a longer period with 127 epochs of spectra over 2013-2022 using the respective broad-line flux, which is a lagged representation of the continuum flux, and broad-line widths. Table \ref{lineBreathingComparison} compares the slopes of the relationship between broad-line width (both FWHM and $\sigma$) and broad-line flux from the present study and from \cite{Wang2020}. We find almost identical results for $\beta_{\sigma}$ (slope for $\sigma$ line-width) for \MgII\ and \Ha\ and $\beta_{\sigma}$ is fairly close for \Hb. However, we see large disparities between $\beta_{\rm FWHM}$ for all three broad emission lines, likely due to the differences in flux (continuum vs. broad-line) used in the analyses.

\begin{figure*}[ht]
\epsscale{1.1}
\plotone{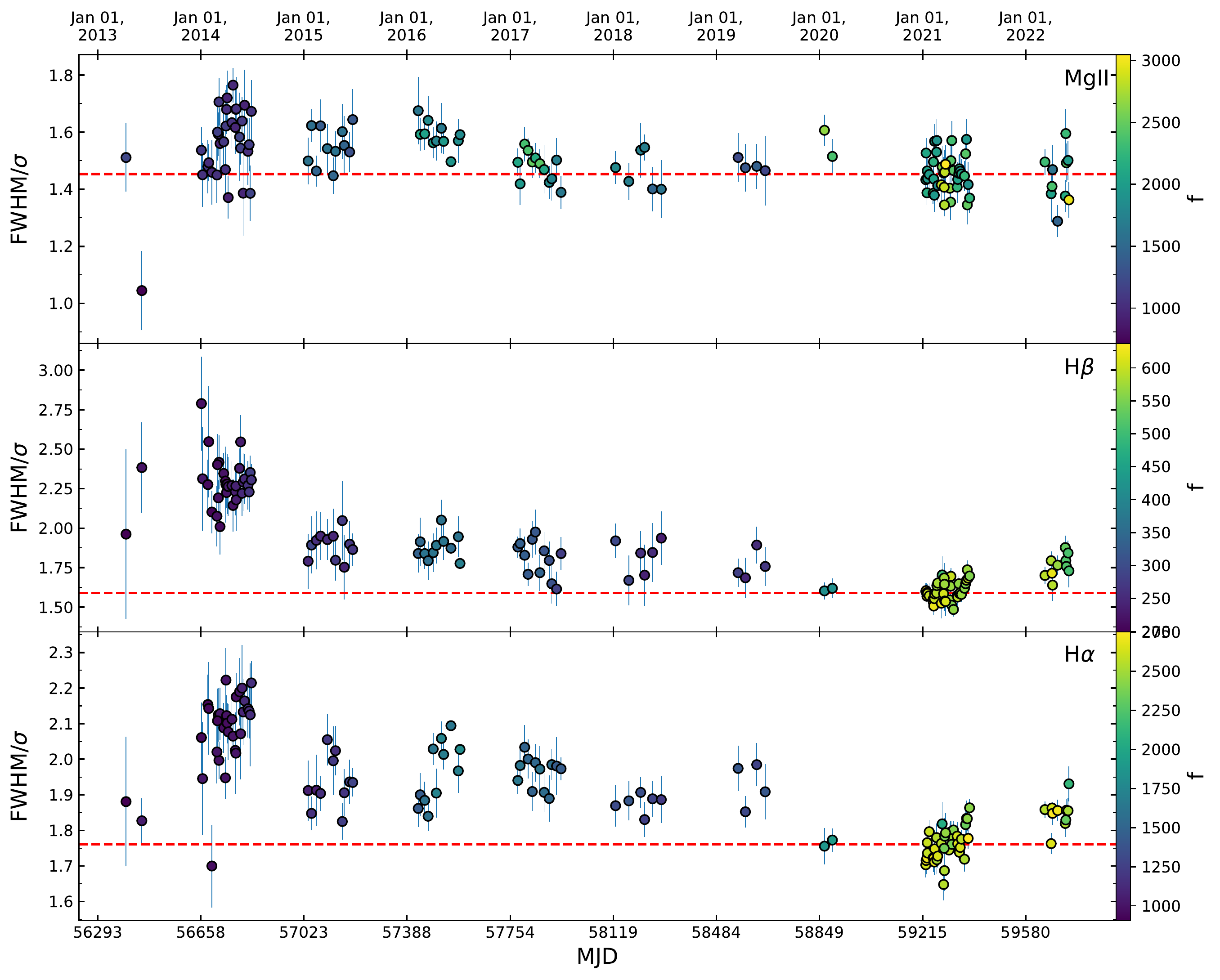}
\figcaption{Ratio of FWHM and $\sigma$ vs. MJD as quantified by our non-parametric measurements of FWHM and $\sigma$ for \MgII\ (top row), \Hb\ (middle row), and \Ha\ (bottom row). The dashed red line indicates the median $\frac{\rm FWHM}{\sigma}$ value for the 2021 monitoring period and the colorbar represents the integrated flux for each emission line in units of $10^{-17}$ erg s$^{-1}$ cm$^{-2}$. The ratio of FWHM and $\sigma$ indicates the peakiness of the distribution. A $\frac{\rm FWHM}{\sigma}$ = 2.355 would indicate a Gaussian distribution, while a value higher than 2.355 would indicate a boxier line and a value lower than 2.355 would indicate a cuspier line. We see that the peakiness of \MgII\ is relatively stable over our 9 year monitoring period, while \Hb\ and \Ha\ become more peakier with time and as they get brighter.
\label{Fig:line_width_ratios}}
\end{figure*}

In Figure~\ref{Fig:NonParaLineBreathing}, there are distinct differences in the line-breathing slopes measured from line-width $\sigma$ and FWHM. We investigate the differences in FWHM and $\sigma$ further in Figure~\ref{Fig:line_width_ratios}, which shows the ratio of FWHM to $\sigma$ as a function of time with a colorbar indicating the relative emission-line flux at each epoch. The FWHM/$\sigma$ ratio measures the boxiness/peakiness of an emission line, with $\frac{\mathrm{FWHM}}{\sigma} = 2.355$ for a Gaussian, a higher ratio ($\frac{\mathrm{FWHM}}{\sigma} > 2.355$) for a boxy line (i.e., low kurtosis), and a lower ratio ($\frac{\mathrm{FWHM}}{\sigma} < 2.355$) for a peaky line (i.e., high kurtosis). The \MgII\ line in \target\ becomes slightly peakier with time (and with increasing flux). The Balmer lines change much more dramatically, going from Gaussian or boxy profiles at a low-flux at the beginning of the monitoring to much peakier profiles at late times and high fluxes.

The change from boxier to peakier broad-line profiles explains the difference in line-breathing slopes measured for FWHM and $\sigma$. The changes in received continuum (as measured in the varying emission-line flux) do not drive monolithic changes in the observed BLR orbits of \target, and instead result in changes to the shape of the velocity distribution. 
This likely indicates asymmetry in the radial distribution of the broad-line gas
% with a higher gas density at lower radii (and higher velocities) observed at low flux that ceases to contribute to the line profile when over-ionized by higher continuum flux. 
% during the luminous phase: we have more 
% \edit1{with the emission-line getting peakier because it gains flux from more distant material.}
where more contribution from distant, low velocity gas increases the line core and makes it appear more peakier in the luminous phase.
We return to a discussion of the complex kinematics of the BLR of \target\ in Section~\ref{LineProfileChanges}.

\begin{table*}[t!]
\centering
\begin{tabular}{c c c c c c c}

 & \multicolumn{2}{c}{\MgII} & \multicolumn{2}{c}{\Hb} & \multicolumn{2}{c}{\Ha} \\
  \cmidrule(lr){2-3} \cmidrule(lr){4-5} \cmidrule(lr){6-7}
Study & $\beta_{\rm FWHM}$ & $\beta_{\sigma}$ & $\beta_{\rm FWHM}$ & $\beta_{\sigma}$ & $\beta_{\rm FWHM}$ & $\beta_{\sigma}$ \\
\toprule
       Wang \etal\ 2020  $\arrvline$ & -0.06 $\pm$ 0.01 & -0.08 $\pm$ 0.01 & -0.08 $\pm$ 0.01 & -0.08 $\pm$ 0.01 & -0.09 $\pm$ 0.01 & -0.16 $\pm$ 0.01 \\
       This work  $\arrvline$ & -0.19 $\pm$ 0.01 & -0.11 $\pm$ 0.01 & -0.47 $\pm$ 0.02 & -0.18 $\pm$ 0.01 & -0.32 $\pm$ 0.01 & -0.17 $\pm$ 0.01\\
\bottomrule
\end{tabular}
\caption{Table comparing the best-fit slopes of the relationship between broad-line width (both FWHM and $\sigma$) and broad-line flux. We note that the results match very well for the \MgII\ and \Ha\ $\sigma$ slopes, but the FWHM slopes for \MgII, \Hb, and \Ha\ have large disparities. The differences between slopes measured by \cite{Wang2020} and this study are likely due to line-profile changes between 2014-2017 and 2018-2022, differences in fitting methods, and the difference in flux (continuum vs. broad-line) used.}
\label{lineBreathingComparison}
\end{table*}

\subsection{The Binary Supermassive Black Hole Hypothesis}
\label{SMBH_binary}

\begin{figure*}[t]%[ht!]
\epsscale{1.1}
\plotone{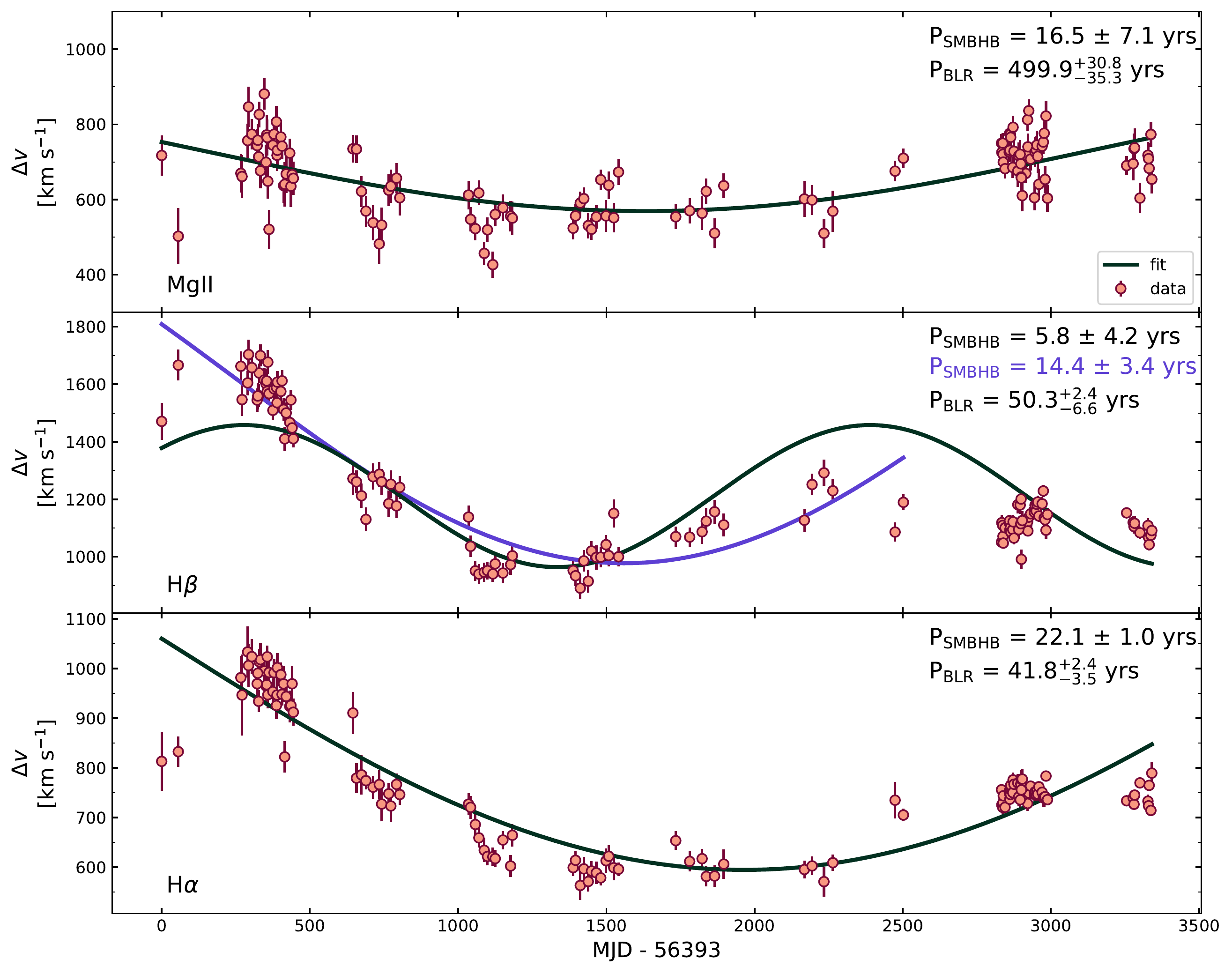}
\figcaption{Sine fits to the radial velocity shifts of \MgII\ (top panel), \Hb\ (middle panel), and \Ha\ (bottom panel). The tangerine points represent the observations and the black line is the best-fit sine function using the fitting routine \texttt{LMFIT} \citep{Newville2014}. The purple line in the middle panel is the best-fit sine function for the \Hb\ radial velocity shifts excluding the 2021 and 2022 data (MJD - 56393 < 2555). For each emission line, we present the period of the best-fit sine function and the period of the BLR at the top right of each panel. We disfavor a binary explanation for the radial velocity variations because the binary period is similar to the broad-line orbital period, and because the three lines have best fit sine functions with different periods.
\label{Fig:sinefitall}}
\end{figure*}

A periodic Doppler shift of the broad emission lines could be indicative of a SMBH binary \citep{Eracleous2012}. However, in the binary scenario, $\Delta{v}$ should be consistent through the different broad emission lines as the systemic velocities should all trace the center of gravity of the active SMBH. This contradicts the observations of \target\ (see: Figure~\ref{Fig:sinefitall}), where $\Delta{v}$ is different for the different broad lines. The differences in $\Delta{v}$ for different lines provide evidence to disfavor a SMBH binary as the cause of the radial velocity variations of \target. The simplest binary scenario (i.e., a scenario in which only one SMBH is active), should also have only bulk radial velocity shifts of the broad emission lines with no variability in their shapes (Fig. 1 in \citealp{Guo2019}), contradicting the ${\rm FWHM}/\sigma$ variations observed in our observations (Figure \ref{Fig:line_width_ratios}).
A binary scenario with velocity shifts driven entirely by periodic Doppler motion would also have a $\Delta{v}$ curve that is symmetric about zero, opposite in shape (concave down) to our observed (concave up) $\Delta{v}$ curves.
% In addition, searches for SMBH binaries typically assume that the broad lines have only bulk radial velocity shifts with no variability in the broad emission line shapes (e.g., Fig. 1 in \citealp{Guo2019}). This would mean that FWHM and $\sigma$ would remain constant throughout which directly contradicts our observations. The variability of the broad-line profiles in \target\ provides further evidence against the SMBH binary hypothesis.

% We note that in the simplest binary scenario (i.e., a scenario in which only one SMBH is active), we would expect to observe only bulk radial velocity shifts of the broad emission lines with no variability in the broad emission line shapes (Fig. 1 in \citealp{Guo2019}). This would mean that FWHM and $\sigma$ would remain constant throughout which directly contradict our observations. 

We provide further evidence against the binary hypothesis by considering
% consider the hypothesis of a SMBH binary starting with
the relationship between binary separation and orbital period (Kepler's third law):

\begin{equation}
    \label{KeplersThird}
    a^{3} = \frac{GM}{4\pi^{2}} P^{2}
\end{equation}
where $a$ is the binary separation, $G$ is the gravitational constant, $M$ is the total binary mass, and $P$ is the period of the binary.

There are two possibilities for a binary SMBH system that has a BLR undergoing radial velocity shifts: a close-pair binary with a circumbinary BLR ($a \ll R_{\rm BLR}$ and $P \ll P_{\rm BLR}$) or a wide-separation binary in which each SMBH has its own BLR ($a \gg R_{\rm BLR}$ and $P \gg P_{\rm BLR}$). A binary SMBH with a separation similar to the BLR ($a \approx R_{\rm BLR}$) would disrupt the orbiting gas and such a system would not have observable broad emission-lines (unlike \target).

In Figure~\ref{Fig:sinefitall}, we fit the radial velocity shifts of all three broad emission lines using a sine function with the \texttt{LMFIT} package \citep{Newville2014}. The best-fit sine functions have characteristic periods of $P(\MgII) = 16.5 \pm 7.1$~yr, $P(\Hb) = 5.8 \pm 4.2$~yr (or $P(\Hb) = 14.4 \pm 3.4$~yr if fit to only the 2014-2020 data), and $P(\Ha) = 22.1 \pm 1.0$~yr. 

Using Equation~\ref{KeplersThird} and assuming that the RM mass ($\log{M_{\rm BH}/M_\odot} = 7.8$) for \target\ is the total binary mass (which is the case for the binary scenario with a circumbinary BLR) and that the radial velocity shifts correspond to the period of the binary, the best-fit periods imply binary separations of $a(\MgII) = 14.9 \pm 8.5$~light-days, $a(\Hb) = 7.4 \pm 6.0$~light-days (or $a(\Hb) = 13.6 \pm 5.2$~light-days if fit to only the 2014-2020 data), and $a(\Ha) = 18.1 \pm 2.3$~light-days. We also use Equation~\ref{KeplersThird} to calculate BLR periods using BLR radii equal to the measured reverberation lags from \citet{Grier2017b} for the Balmer lines and from \citet{Homayouni2020} for the \MgII\ line. These periods are shown on the top right of each panel in Figure~\ref{Fig:sinefitall}. 
%% YS: clarify that you use the RM BH mass for the total binary mass -- btw this is not entirely self-consistent, since the RM mass assumes the BLR gas is on virial orbits around a single BH)}
% Figure~\ref{Fig:sinefitall} also includes BLR periods calculated from Equation~\ref{KeplersThird} using BLR radii equal to the measured reverberation lags from \citet{Grier2017b} for the Balmer lines and from \citet{Homayouni2020} for the \MgII\ line.

The best-fit sine periods range from 6 to 22~yr, with semi-major axes of 7 to 18~light-days for the implied binary orbit. Comparing these values to the observed lags from \cite{Grier2017b} and \cite{Homayouni2020}, we find that the range of binary semi-major axis is similar to the observed lag for \Ha\ (i.e., $a_{\rm BLR, H\alpha} = 27.7^{+5.3}_{-4.7}$ light-days). In the binary scenario this would place the black hole orbit within the BLR, which would cause the BLR to become unstable due to the gravitational interactions with the orbiting black hole. We therefore disfavor a binary explanation for the radial velocity shifts in \target.

If we instead consider the possibility of a wide separation binary system with a BLR around one (active) black hole, then the RM mass represents a minimum for the total binary mass and the semi-major axes from Equation~\ref{KeplersThird} are also minimum values. Even for a maximum total binary mass of $5 \times 10^{10} M_{\odot}$ (the upper limit for an AGN, e.g., \citealt{King2016}), and an extreme binary mass ratio of $\sim$1000:1, the binary separation would be $a(\rm \MgII) = 166.8 \pm 5.1$~light-days, again overlapping with the observed BLR size of $a_{\rm MgII} = 106$~light-days \citep{Homayouni2020}. In other words, the binary hypothesis is ruled out for both a circumbinary BLR and a single-AGN BLR due to overlap between the putative binary orbit and the observed BLR size.

\subsection{BLR Kinematics and Geometry}
\label{LineProfileChanges}

We begin by calculating the dynamical timescale for the BLR, $\tau_{\rm dyn} \simeq (R_{\rm BLR}^{3} / G M_{\rm BH})^{1/2}$ = 7.7 yr, using the \Hb\ based parameters for $R_{\rm BLR}$ and $M_{\rm BH}$. The best-fit variability periods in Figure~\ref{Fig:sinefitall} are longer than this dynamical timescale, indicating that the observed variability is consistent with dynamical changes in the BLR. We present an example of a phenomenological model for these dynamical changes below.
% (first, let's review)
% \target\ has large ($\sim$300-700~km~s$^{-1}$) radial velocity shifts in its broad emission lines. Although the source exhibits line breathing in a fashion that is similar to other quasars, its line shifts do not follow the same pattern as the changes in line brightness and width.
 
% As established in the previous subsection, the radial velocity shifts are best-fit with periods that are similar to the BLR orbital period computed from the previous reverberation lags of the Balmer lines. This rules out a binary SMBH as the cause of the line shifts, and instead suggests that the radial velocity variations are connected to BLR orbits.

% (now explain the inflow part)
% Our preferred
\target\ has large ($\sim$400-800~km~s$^{-1}$) radial velocity shifts in its broad emission lines. Although the source exhibits line breathing in a fashion that is similar to other quasars, its line shifts do not follow the same pattern as the changes in line brightness and width. One plausible explanation of the radial velocity shifts is a BLR with azimuthal asymmetry and a gradient of inflow velocity in the radial direction. An inflow model is motivated by previous velocity-resolved reverberation mapping observations  \citep[e.g.,][]{Bentz2010b, Grier2017a, Bentz2021, U2022, Villafana2022} that frequently find evidence for inflowing BLRs in nearby Seyfert AGN. Alternatively, we could be seeing an outflow due to BLR emission being preferentially emitted back toward the ionizing source \citep{Ferland1992}, with a gradient of outflow velocity decreasing with radial distance from the quasar. We refer to an inflow hereafter, but acknowledge that a decelerating outflow with the right azimuthal asymmetry might also explain the observations.

\begin{figure*}[h]
\epsscale{0.9}
\plotone{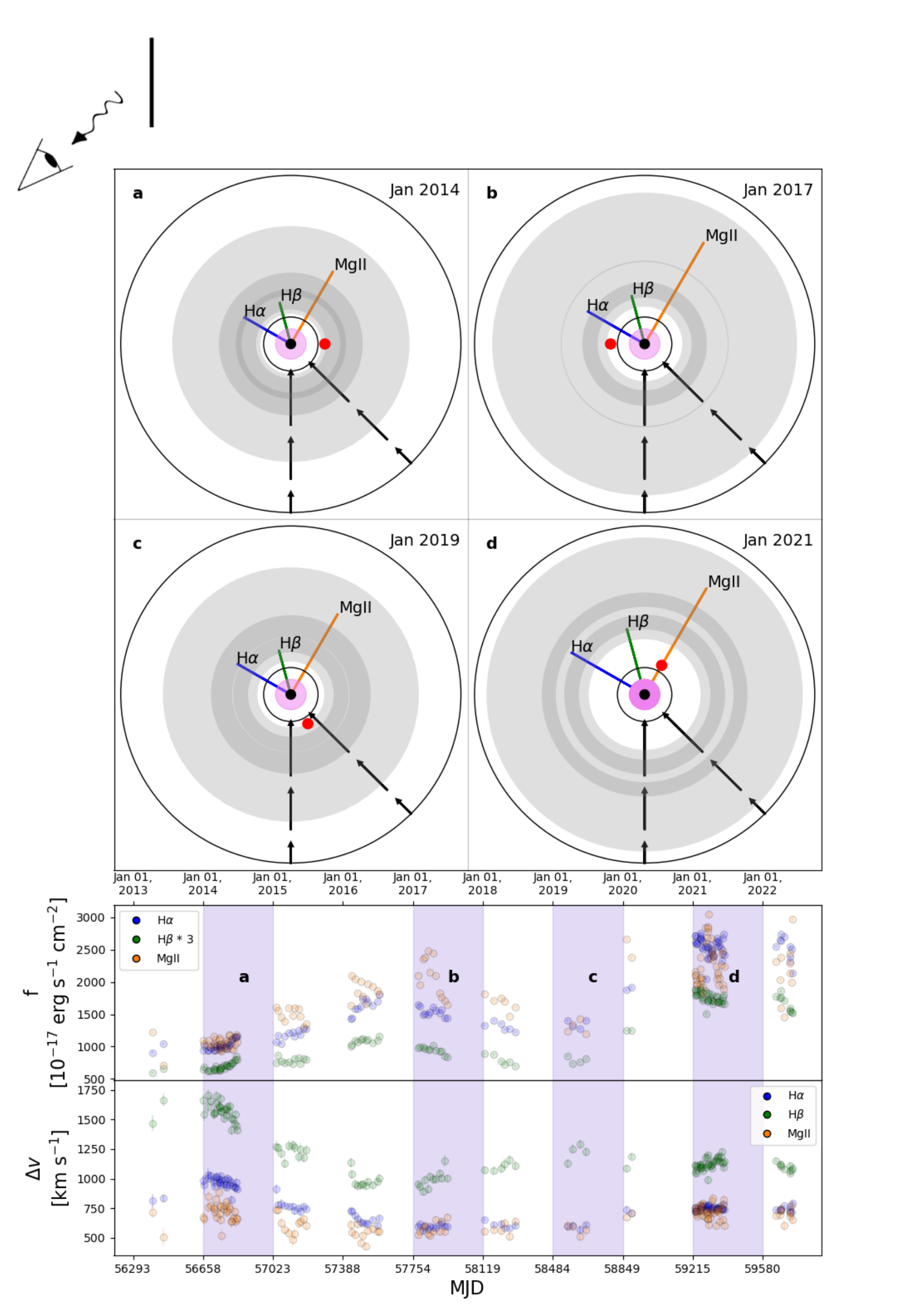}
\figcaption{
Our geometric and kinematic model of the BLR in \target\ (top), connected to the observed broad-line flux light curve and radial velocity shifts (bottom). The viewing angle of the observer is indicated in the top left. The radii of \Hb, \Ha, and \MgII\ are indicated by green blue, and orange in both the top and bottom panels, scaled to match the measured reverberation lags \citep{Grier2017b,Homayouni2020} and with gray shading to indicate the radial extent of each emission line region. The broad-line regions ``breathe,'' moving inward when the quasar is fainter (e.g., in 2014 and 2019) and outward when the quasar is brighter (e.g., in 2021). The red dot represents an azimuthal asymmetry in the BLR that orbits the black hole during our spectroscopic monitoring. In the animation, darker circles indicate the flux and radial velocity in the bottom panels at the same time as the model in the top panels. In the still images, the labels a, b, c, and d indicate representative time windows (also shaded purple) for the BLR model in the top panels and corresponding flux and line-center in the bottom panels. A combination of radial inflow (vectors), azimuthal asymmetry (red spot), and line breathing (green, blue, and red lines and gray shading) can explain the unusual broad-line variability of \target. 
% The online version of this Figure is an animation that shows our phenomenological model of the BLR throughout the course of our monitoring.
The online version of this figure is an animation. The animation is 0:24 minutes long and shows the time-evolution of our BLR model from 2013 to 2022 (top) along with the corresponding time-evolution of the flux and line-center, respectively, in the bottom panels.}
\label{Fig:kinematicmodel}
\end{figure*}

Our model is illustrated by Figure \ref{Fig:kinematicmodel}, which includes an animation of the BLR and its kinematics connected to the observed broad-line light curves and widths.

We begin by noting that all three emission lines have large redshifts ($\sim$500-1600~km~s$^{-1}$) with respect to the narrow lines (i.e., the systemic redshift) at all epochs. This likely indicates bulk \textit{inflow} of the BLR gas in the line of sight. At all epochs the \Hb\ line is most redshifted ($\sim$900-1600~km~s$^{-1}$) and the \MgII\ line is the least redshifted ($\sim$500-800~km~s$^{-1}$). The reverberation lags indicate that the \MgII-emitting gas (rest-frame $R_{\rm BLR} = 106$~light-days; \citealp{Homayouni2020}) is much further from the continuum emission than the \Hb-emitting gas (rest-frame $R_{\rm BLR} = 23$~days; \citealp{Grier2017b}). This further implies that the inflow of the BLR gas has a radial gradient, with higher inflow velocity for gas closer to the SMBH (like the \Hb\ emission region) and lower inflow for more distant gas (like the \MgII\ emission region). The \Ha\ line is a bit puzzling in this picture because it has a similar reverberation lag (rest-frame $R_{\rm BLR} = 20$~days; \citealp{Grier2017b}) to \Hb\ but has a smaller inflow velocity. In general the \Ha\ line is expected to be emitted from slightly larger radii than \Hb\ due to radial stratification and optical depth effects \citep{Netzer1975, Rees1989, Korista2004, Bentz2010a} and we assume that this is also the case here, despite the similarity in measured reverberation lags for the two lines.

Alternatively, the redshifted broad emission lines might be explained by gravitational redshift \citep{Tremaine2014}. This scenario similarly predicts that the \Hb\ line would be more redshifted than the \Ha\ and \MgII\ lines, due to \Hb\ being emitted from gas closer to the black hole that has larger orbital velocities. However the broad emission line widths ($\sigma<3000$~km~s$^{-1}$ for all three lines) do not imply relativistic orbits unless the BLR is observed at a nearly face-on inclination. We thus prefer an inflow as the explanation for the redshifted broad emission lines rather than gravitational redshift.
%% (YS: Gravitational redshift would also be fixed, but the centroids of the lines are changing.)

% (now the asymmetry)
An inflowing BLR explains the redshifted lines, and a radial gradient inflow explains the difference in redshift from \Hb\ to \Ha\ to \MgII. But it does not explain the \textit{variability} of the broad-line centers. Figure~\ref{Fig:sinefitall} shows that the observed line-center variations, especially for \Ha\ and \MgII, are best-fit by a sine function with a period that is similar to the BLR orbital period implied from the \Hb\ lag (i.e., 22.1 years for \Ha\ and 16.5 years for \MgII). This suggests that the line-center shifts might be related to an azimuthal asymmetry in the broad-line emission that orbits the central SMBH. The asymmetric BLR emission might be associated with a higher density in the gas, a hot spot (or hot ``wedge'' or spiral arm), higher responsivity of the gas on one side, or could be associated with asymmetric illumination from the accretion disk.

% (now connect inflow kinematics + asymmetric geometry)
The detailed radial velocity shifts can be explained by a combination of inflowing gas onto the BLR with a radial gradient, orbiting asymmetric gas emission, and flux-driven changes to the optimal emission region (line breathing). At the start of our monitoring in $\sim$2014, the BLR receives low continuum flux and the emission region is close to the SMBH, with high bulk inflow velocity and an asymmetric gas region that is additionally on the receding (redshifted) side of its orbit. The asymmetric gas region reaches the approaching part of the orbit in $\sim$2017, although the line is still redshifted due to the bulk inflow. The modest ($\sim$2$\times$) brightening of the line emission and coordinated decrease in line-width over 2014-2017 causes the line emission region to move slightly outward, also resulting in slightly lower redshift of the line. After 2017 the asymmetric region of the BLR begins receding again, and together with the line flux decrease (and line width increase, and emission radius decrease) the line once again becomes redder. We hypothesize that, in $\sim$2020, the orbit of the asymmetric region would have caused the lines to reach the high redshifts they began with in 2014, if not for the dramatic brightening of the Balmer lines observed in 2020. The significant increase in flux results in the line emission region becoming larger (with observed lower line widths), corresponding to lower bulk inflow velocities. Combined with the redder emission from the orbiting azimuthal asymmetry, the decreased bulk inflow velocity results in a lower integrated redshift velocity for the line in 2020 as compared to 2014.

%% [YS: here is a good place to rule out the asymmetry in the velocity-resolved transfer function as the cause of the centroid shift. Barth et al. argued that if the BLR is an inflowing geometry, the red wing would respond earlier than the blue wing. So when the continuum flux is increasing, we'd expect more redshifted line centroid, which is opposite to what you observe in Fig. 11. So reverberation mapping effects (from an asymmetric transfer function) are insufficient to explain the observed centroid shift.]

The combined effects of bulk inflow, azimuthal asymmetry, and line breathing are strongest for \Hb, as the line is at a smaller $R_{\rm BLR}$, and weakest for \MgII, as the line is at a much larger $R_{\rm BLR}$. The pattern over 2014-2020 of the \MgII\ line-center shifts are much more symmetric than the \Hb\ line-center shifts. The \Hb\ line also has much larger redshift velocities than the other lines due to its emission region occupying smaller BLR radii.

% An alternative explanation to inflow is the reddening of the broad emission lines due to gravitational redshift \citep{Tremaine2014}. In this scenario, there would also be a gradient where emission lines that are closer to the SMBH (i.e., \Hb) are more red than the ones that are further away (i.e., \MgII). The \Hb\ line is likely affected by gravitational redshift apparent in the red wing that appears (For example, see: Figures~\ref{Fig:GIFexample},~\ref{Fig:CompositeSpectra}) For quasars with broad emission lines that are this narrow ($\sigma \sim$~2000~km~s$^{-1}$), and at a modest inclination, general relativistic effects are negligible ($\gamma \sim$ 1). However in an extreme face-on scenario (I $\sim$ 15), this kind of inclination could have strong gravitational redshifts since in an extreme face-on inclination the velocity of the BLR would be minuscule and there needs to be a mechanism to broaden the observed emission lines.

% Now end the subsection with a paragraph to note that the next step is velocity-resolved RM.
Velocity-resolved reverberation mapping would provide a test of our geometric and kinematic model for the BLR in \target\ shown in Figure \ref{Fig:kinematicmodel}. Velocity-resolved RM
% In order to distinguish the cause of the systemic reddening of the emission lines (inflow vs. gravitational redshift), we need to do velocity resolved reverberation mapping. This
involves measuring how different segments of an emission line reverberate in response to continuum variations \citep{Denney2009, Bentz2009, Bentz2010a, U2022, Li2022, Villafana2022}. If the BLR is virialized, the lags at the center of the emission line are the longest since they preferentially correspond to gas further from the black hole, with shorter lags measured for the line wings. For an inflowing BLR (like our phenomenological model), the lags of the blue wing would be longest and the lags of the red wing would be shortest.
%and for an outflowing model, the lags of the red wing are the longest and the lags of the blue wing are the shortest.
%Velocity resolved reverberation mapping will help us determine if there is in fact a bulk inflow in the BLR. 
Velocity-resolved RM could also isolate the putative azimuthal asymmetry that orbits around the black hole in our model. We anticipate performing velocity-resolved RM and further testing our model for \target\ in future work.

\section{Summary}
\label{Sec5}
We presented multi-epoch optical spectroscopy of \target, a luminous quasar which exhibits unusual broad emission-line variability in the SDSS-RM field. This object was identified from a broad search for extreme variability in quasar broad emission-line profiles and has been observed 127 times over the 9 year monitoring period with plans to continue observations through 2026, within the SDSS-V project. 

We find that \target\ exhibits normal line-breathing behavior consistent with many previously studied AGN (e.g., \citealt{Barth2015} and \citealt{Wang2020}) in the variations of flux and line-width $\sigma$, but there is a sizable discrepancy between line-width $\sigma$ and FWHM for the Balmer series (\Hb\ and \Ha). We find that the shape of the line-profile (as indicated by FWHM/$\sigma$) changes over time for the Balmer series from a boxy line profile (FWHM/$\sigma$ > 2.355) to a cuspy profile (FWHM/$\sigma$ < 2.355) with increasing flux throughout the 9 year monitoring period. This likely indicates asymmetry in the radial distribution of the broad-line gas, where more contribution from distant, low-velocity gas increases the line core and makes it appear more peakier in the luminous phase.
% with a higher gas density at lower radii (and higher velocities) observed at lower flux that ceases to contribute to the line profile when over-ionized by higher continuum flux.

% We found dramatic radial velocity shifts
Dramatic radial velocity variations occur
in each of the three broad emission lines (\MgII, \Hb, and \Ha) that all follow the same qualitative trend of starting red, shifting bluer over a few years, and then getting redder near the end of the monitoring. 
% 
% We explored the possibility of the radial velocity shifts being caused by a supermassive black hole binary system. We found that the characteristic period, assuming a binary, is approximately the same as the period of the BLR. This would put the orbiting black hole directly inside of the BLR and this would gravitationally disrupt the BLR and cause it to be unstable, which would not allow \target\ to have observed broad emission-lines. We also note that the sine fits look different for different emission lines, which suggests that the radial velocity shifts are not well-described by a sine function. The binary semi-major axis being similar to the observed lag for \Ha, and the poor fit of a periodic sinusoidal function provide us reasons to disfavor a supermassive black hole binary as the cause of the radial velocity variations.
The radial velocity shifts are not well explained by a black hole binary because the best-fit period corresponds to the inner BLR orbits, such that a putative binary would have disrupted the BLR gas.

Our explanation for the large radial velocity shifts in the broad emission lines of \target\ is a BLR with azimuthal asymmetry and a gradient of inflow velocity in the radial direction coupled with
%the fact that there are
flux-driven changes to the optimal emission region (line breathing), as illustrated in Figure \ref{Fig:kinematicmodel}.
Similar instances of line-profile variability due to complex gas kinematics in the BLR are likely to represent an important source of false positives in radial velocity searches for binary black holes. The long-duration, wide-field, and many-epoch spectroscopic monitoring of SDSS-V/BHM-RM will be excellent for studying such systems and helping understand the various mechanisms driving BLR dynamics.

\section{Acknowledgements}
LBF, JRT, and MCD acknowledge support from NSF grant CAREER-1945546, and with CJG acknowledge support from NSF grant AST-2108668. JRT, CJG, and YS also acknowledge support from NSF grant AST-2009539. MK acknowledges support by DFG grant KR 3338/4-1.
B.T. acknowledges support from the European Research Council (ERC) under the European Union's Horizon 2020 research and innovation program (grant agreement 950533) and from the Israel Science Foundation (grant 1849/19). X.L. acknowledges support from NSF grant AST-2206499. CR acknowledges support from the Fondecyt Iniciaci\'on grant 11190831 and ANID BASAL project FB210003. RJA acknowledges support by ANID BASAL FB 210003 and by FONDECYT grant number 1191124. M. L. M.-A. acknowledges financial support from Millenium Nucleus NCN$19\_058$ (TITANs).

Funding for the Sloan Digital Sky Survey V has been provided by the Alfred P. Sloan Foundation, the Heising-Simons Foundation, the National Science Foundation, and the Participating Institutions. SDSS acknowledges support and resources from the Center for High-Performance Computing at the University of Utah. The SDSS web site is \url{www.sdss5.org}.

SDSS is managed by the Astrophysical Research Consortium for the Participating Institutions of the SDSS Collaboration, including the Carnegie Institution for Science, Chilean National Time Allocation Committee (CNTAC) ratified researchers, the Gotham Participation Group, Harvard University, Heidelberg University, The Johns Hopkins University, L'Ecole polytechnique f\'{e}d\'{e}rale de Lausanne (EPFL), Leibniz-Institut f\"{u}r Astrophysik Potsdam (AIP), Max-Planck-Institut f\"{u}r Astronomie (MPIA Heidelberg), Max-Planck-Institut f\"{u}r Extraterrestrische Physik (MPE), Nanjing University, National Astronomical Observatories of China (NAOC), New Mexico State University, The Ohio State University, Pennsylvania State University, Smithsonian Astrophysical Observatory, Space Telescope Science Institute (STScI), the Stellar Astrophysics Participation Group, Universidad Nacional Aut\'{o}noma de M\'{e}xico, University of Arizona, University of Colorado Boulder, University of Illinois at Urbana-Champaign, University of Toronto, University of Utah, University of Virginia, Yale University, and Yunnan University.

\software{{\tt AstroPy} \citep{Astropy2013, Astropy2018},
{\tt Matplotlib} \citep{Matplotlib2007}, 
{\tt NumPy} \citep{NumPy2020}, {\tt SciPy} \citep{scipy2020}}, {\tt linmix} \citep{Kelly2007}, {\tt PyQSOFit} \citep{Guo2018, Shen2019b}, {\tt specutils} \citep{Earl2022}

\appendix{}

\section{Alternative Fitting Procedures}
\label{AltFitting}
Our paper uses a non-parametric approach to measuring the broad emission lines of \target. In this Appendix, we fit the emission-line properties with alternative methods in order to ensure that our conclusions about the quasar's dramatic changes in line flux, width and radial velocity are robust to the choice of fitting method. Section~\ref{SingleGauss} describes the results of single-Gaussian fits for the broad emission lines. Section~\ref{QSOFit} instead uses \texttt{PyQSOFit} \citep{Guo2018, Shen2019b}, a spectral fitting code that fits the quasar continuum, iron psuedo-continuum, and broad and narrow emission lines with multiple Gaussians. Section~\ref{Fit_Method_Comparison} compares the three fitting methods and concludes that our conclusions about the quasar's dramatic broad-line breathing and radial velocity shifts are robust to different fitting methods.

\subsection{Single Gaussian Fits}
\label{SingleGauss}
The first alternative fitting method was to fit the broad emission lines with a single Gaussian. Similarly to our non-parametric measurements, we begin with the same continuum-subtracted spectra describe in Section~\ref{continuum_fits}. We then used the Gaussian1D method from the \textsc{AstroPy} package \citep{Astropy2013, Astropy2018} in order to fit a single Gaussian to each of the narrow and broad emission lines.

\begin{figure*}[t]%[ht!]
\epsscale{1.1}
\plotone{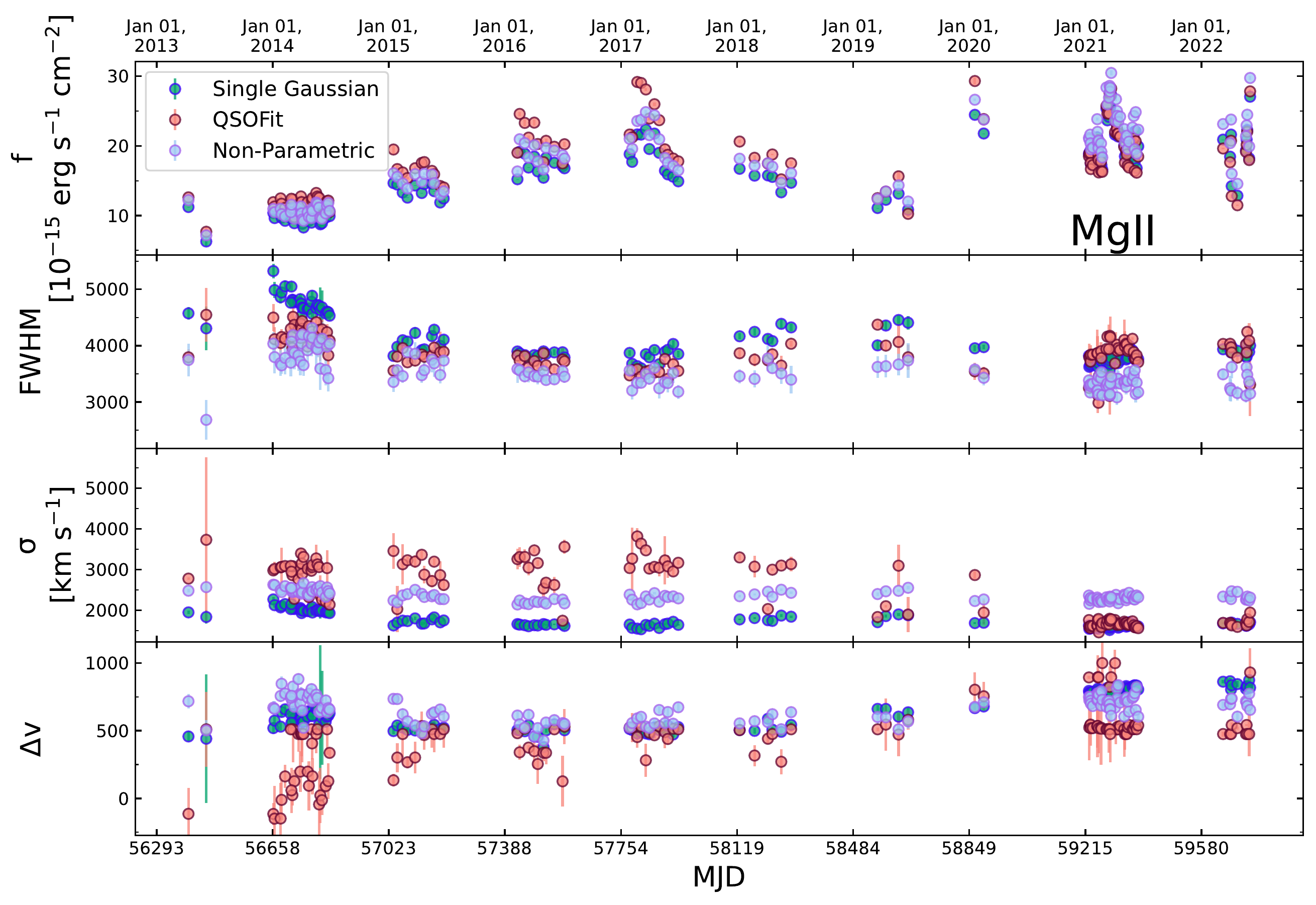}
\figcaption{Variability of the \MgII\ broad emission-line profile, as quantified by our three fitting methods for the line flux (top row), FWHM (2nd row), line-width $\sigma$ (3rd row), and the line-center velocity shifts $\Delta{v}$ (bottom row). The units for FWHM, line-width $\sigma$, and line-center velocity shift $\Delta{v}$ are all in km~s$^{-1}$. The measured flux and FWHM are consistent across all methods. There are differences in the measured $\sigma$ and $\Delta{v}$ due to differences in fitting the wings of the line profile, but all three methods find similar qualitative trends of line breathing and shifts in the line center.
\label{Fig:totalMgIILineProfile}}
\end{figure*}

\begin{figure*}[t]%[ht!]
\epsscale{1.1}
\plotone{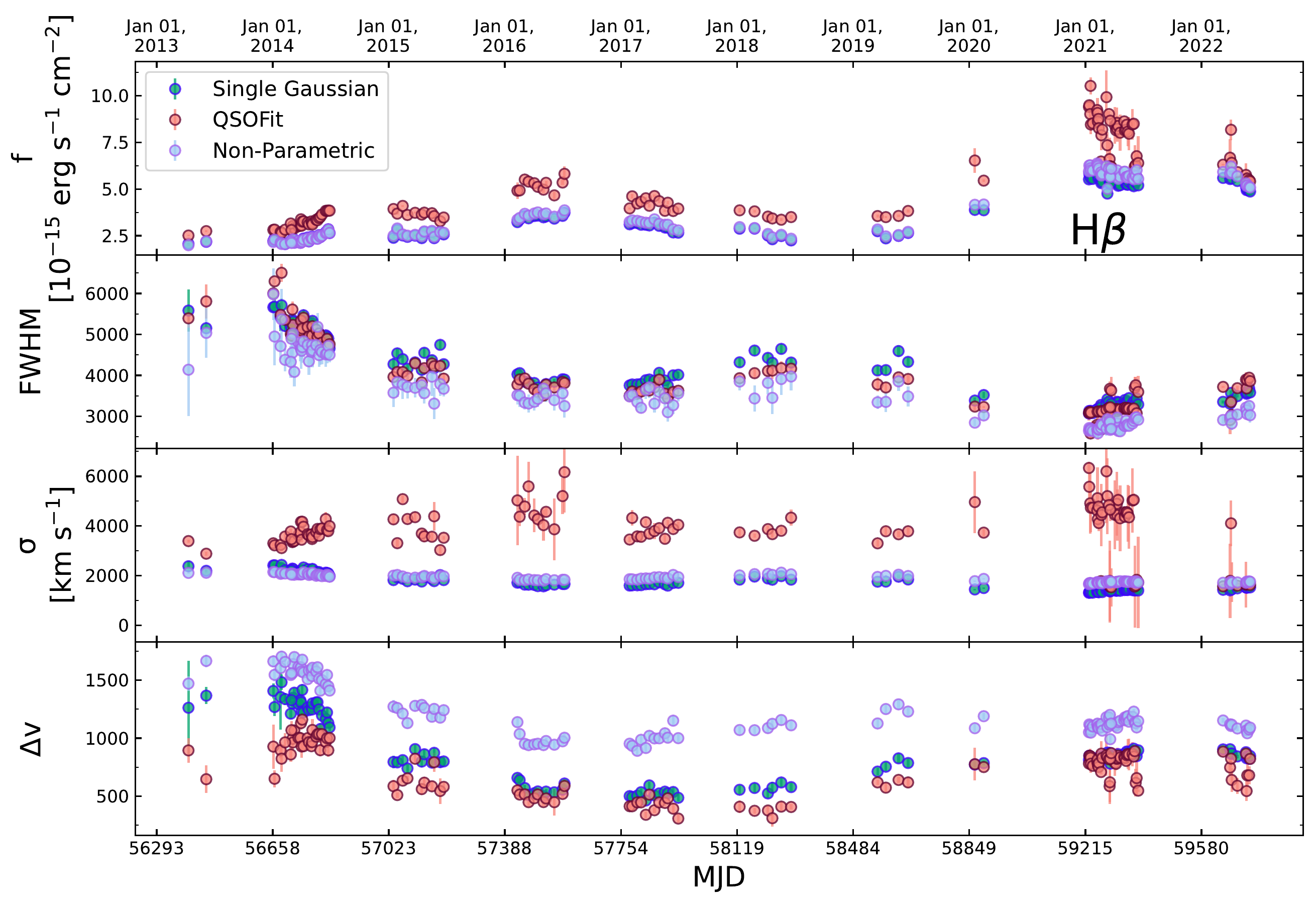}
\figcaption{Variability of the \Hb\ broad emission-line profile, as quantified by the three fitting methods for the line flux (top row), FWHM (2nd row), line-width $\sigma$ (3rd row), and the line-center velocity shifts $\Delta{v}$ (bottom row). The units for FWHM, line-width $\sigma$, and line-center velocity shift $\Delta{v}$ are all in km~s$^{-1}$. The \Hb\ line has a complex profile (see Figure~\ref{Fig:GIFexample}) and the three methods result in different line fluxes, $\sigma$ widths, and line centers due to including different amounts of the line wings in the fit. Despite these differences, all three methods find similar qualitative trends for the variability of the line.
\label{Fig:totalHbLineProfile}}
\end{figure*}

\begin{figure*}[t]%[ht!]
\epsscale{1.1}
\plotone{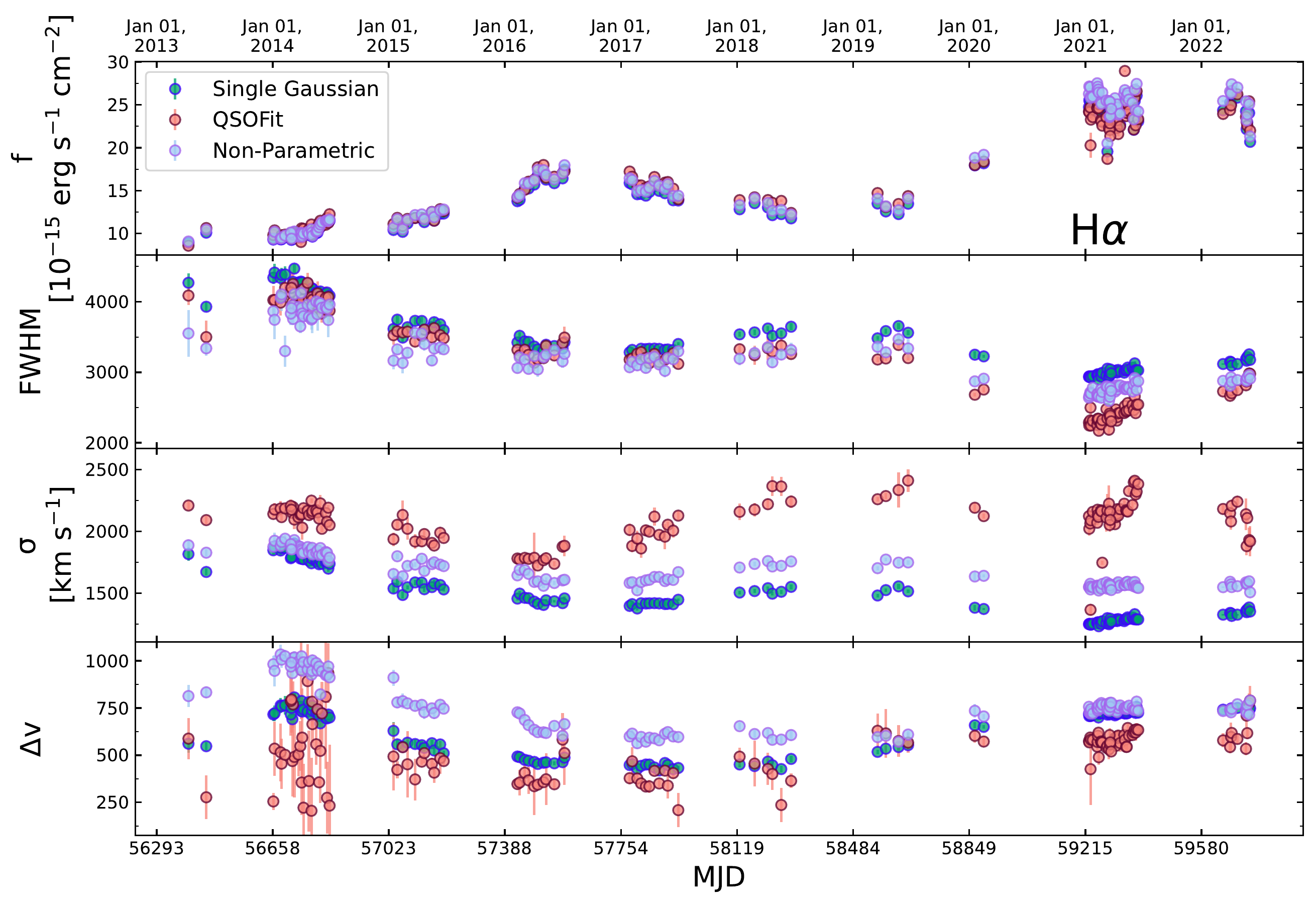}
\figcaption{Variability of the \Ha\ broad emission-line profile, as quantified by our three fitting methods for the line flux (top row), FWHM (2nd row), line-width $\sigma$ (3rd row), and the line-center velocity shifts $\Delta{v}$ (bottom row). The units for FWHM, line-width $\sigma$, and line-center velocity shift $\Delta{v}$ are all in km~s$^{-1}$. As for the fit to \MgII, the measured flux and FWHM is consistent across all methods. The measured flux and FWHM are consistent across all methods.There are differences in the measured $\sigma$ and $\Delta{v}$ due to differences in fitting the wings of the line profile, but all three methods find similar qualitative trends of line breathing and shifts in the line center.
\label{Fig:totalHaLineProfile}}
\end{figure*}

For each individual emission-line region, we followed the initial procedure in Section~\ref{narrow_fits}, in that we tied the line-centers and line-widths of the narrow emission lines to $\OIII\lambda$5007 and constrained the amplitude of $\OIII\lambda$4959 to be 1/3 that of $\OIII\lambda$5007. In what follows, we briefly describe the steps to obtain the broad emission-line component in each emission-line region.

For \MgII, we fit the individual spectra with a single Gaussian as there are no narrow emission lines within the \MgII\ emission-line region. 

\begin{figure*}[ht]
\epsscale{1.1}
\plotone{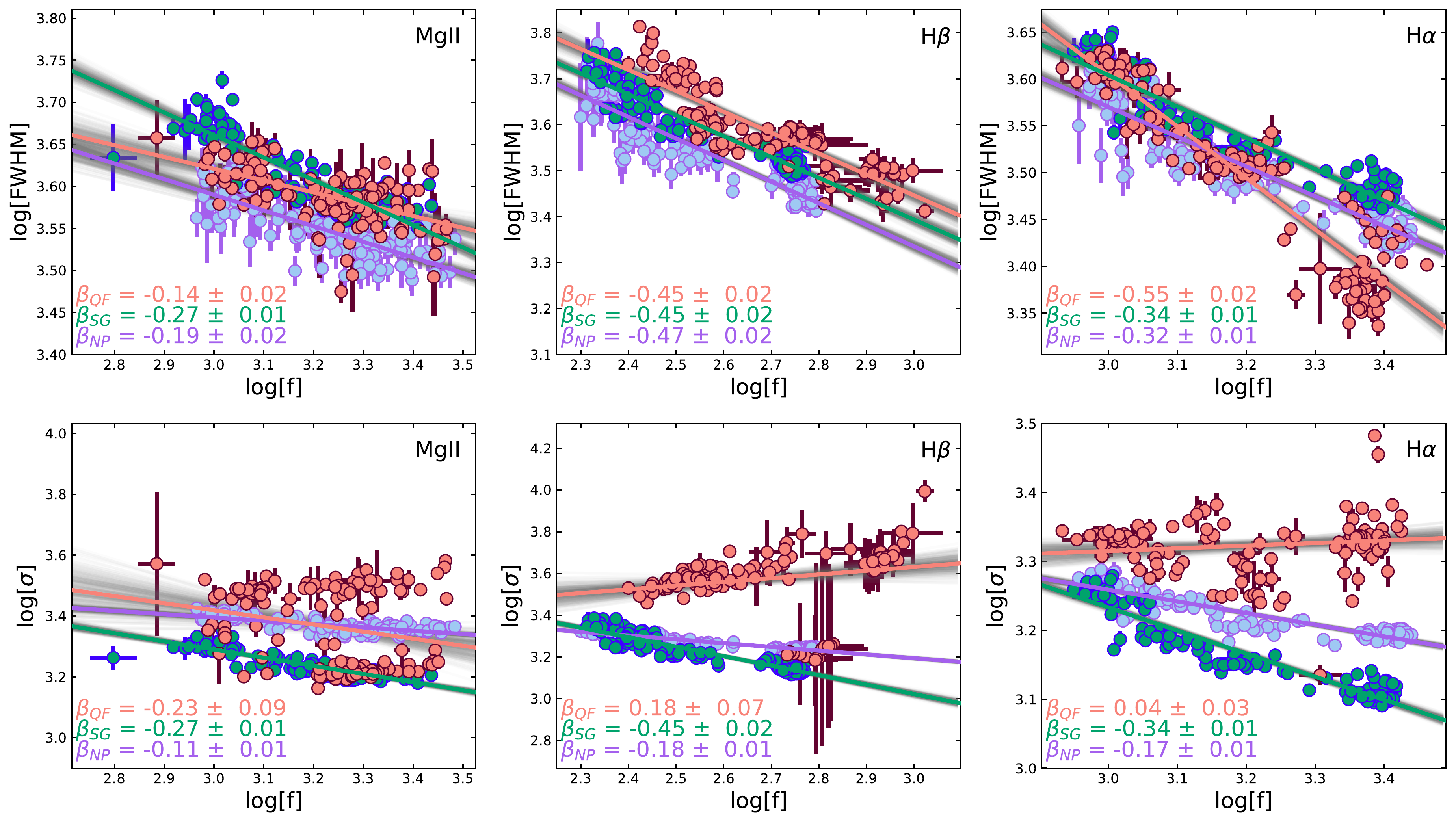}
\figcaption{Line-width vs. flux plots for \target\ for all fitting methods. The purple points are from the non-parametric fitting method, the pink points are from QSOfit, and the green points are from the single Gaussian fitting method. The colors of the lines of best-fit correspond to the colors of the data points. The top panels show the FWHM vs. flux for \MgII, \Hb, and \Ha\ (left to right). The bottom panels show the line-width $\sigma$ vs. flux for \MgII, \Hb, and \Ha\ (left to right). The individual slopes are shown at the bottom left of each panel where $\beta_{QF}$ is the slope for the QSOFit fitting method, $\beta_{SG}$ is the slope for the single Gaussian fiting method, and $\beta_{NP}$ is the slope for our non-parametric measurements. The best-fit slopes for FWHM are broadly consistent, while we see large discrepancies between the slopes for line-width $\sigma$. This seems consistent with what we see for the FWHM values and line-width $\sigma$ values in Figures~\ref{Fig:totalMgIILineProfile}, \ref{Fig:totalHbLineProfile}, and \ref{Fig:totalHaLineProfile} where we see a general consensus for FWHM values across all schemes, and differences between values in line-width $\sigma$.
\label{Fig:LineBreathingAll}}
\end{figure*}

For \Ha, we simultaneously fit the narrow-lines ($\NII\lambda$6548, narrow \Ha, $\NII\lambda$6584, $\SII\lambda$6718, and $\SII\lambda$6732) and the broad \Ha\ emission line with no constraints on the amplitude of any emission line. This simultaneous fitting of all of the narrow lines and the broad \Ha\ component produced reliable fits.

For \Hb, degeneracies between the lines made it difficult to simultaneously fit the narrow lines (narrow \Hb, $\OIII\lambda$4959, and $\OIII\lambda$5007) while also fitting the broad \Hb\ component reliably well. Instead, we constrained the amplitude of the narrow \Hb\ line 
%using the narrow \Hb\ amplitude
to that which produced the smoothest residual in the median spectrum fit. We then applied that narrow \Hb\ amplitude to the fits for all epochs along with the $\OIII\lambda$4959 and $\OIII\lambda$5007 narrow lines and produced a narrow line subtracted spectrum. From that, we fit a single Gaussian to the continuum and narrow-line subtracted spectra in all epochs to fit the broad \Hb\ component.

For all emission lines in this fitting scheme, we calculated the FWHM using the relationship between FWHM and line-width ($\sigma$) for a Gaussian, namely, FWHM = 2.355$\sigma$.

We employed bootstrap resampling to estimate the uncertainties in line flux, line center, line width ($\sigma$), and FWHM similar to our uncertainty estimation approach in Section~\ref{non_parametric}. We resampled each individual spectra 200 times by sampling the fluxes within their Gaussian uncertainties in the respective epoch. We adopted the standard deviations of the 200 re-sampled spectra as our uncertainties for each epoch.

\subsection{PyQSOFit}
\label{QSOFit}
The second alternative fitting method was to fit the entire spectrum with the robust quasar-fitting algorithm \texttt{PyQSOFit}. The considerations for the continuum fit include a power law, an optical and UV FeII template, and a fifth-order polynomial to account for possible dust reddening. \texttt{PyQSOFit} allows the user to select the emission-line and continuum components that are included in the overall fit as well as selecting the range of fitting for the emission lines. We show our fitting parameters in Table~\ref{QSOfitParams}. 

We left most of the default parameters unchanged, except for turning off the dereddening and host decomposition since our fits within the relatively narrow wavelength regions around the broad lines will be unaffected by reddening and contributions from the host galaxy. We decided to fit the narrow emission lines with a single Gaussian, consistent with the fitting methods in Section~\ref{non_parametric} and Appendix~\ref{SingleGauss}. We allowed 2 Gaussians to fit the broad lines in order to better capture the complex shapes of the line profiles.

\begin{table}[h!]
\centering
\begin{tabular}{llcc}
\toprule
        line &   type &  n\_gauss &  $\lambda_{\rm rest}$ \\
\midrule
        \Ha &  broad &      2.0 &          6564.61 \\
        \Ha & narrow &      1.0 &          6564.61 \\
        \NII$\lambda$6549 & narrow &      1.0 &          6549.85 \\
        \NII$\lambda$6585 & narrow &      1.0 &          6585.28 \\
        \SII$\lambda$6718 & narrow &      1.0 &          6718.29 \\
        \SII$\lambda$6732 & narrow &      1.0 &          6732.67 \\
        \Hb &  broad &      2.0 &          4862.68 \\
        \Hb & narrow &      1.0 &          4862.68 \\
        \OIII$\lambda$4959 & narrow &      1.0 &          4960.30 \\
        \OIII$\lambda$5007 & narrow &      1.0 &          5008.24 \\
        \MgII &  broad &      2.0 &          2798.75 \\
         \MgII & narrow &      1.0 &          2798.75 \\
\bottomrule
\end{tabular}
\caption{Table of our parameters used for \texttt{PyQSOFit}. The first column is the emission-line name, the second column is the type of emission line (broad or narrow), the third column is the number of Gaussians used in the fit, and the fourth column is the rest-frame central wavelength.}
\label{QSOfitParams}
\end{table}

\texttt{PyQSOFit} was used to obtain the flux, line width ($\sigma$), FWHM, and line center of the \MgII, \Hb, and \Ha\ emission lines for each epoch. WE use the uncertainties reported by \texttt{PyQSOFit} for each of the fitted quantities.

Note that our \texttt{PyQSOFit} analysis is not exactly analogous to \cite{Wang2020} because they used 3 Gaussians to fit each broad emission line. Furthermore, they rejected epochs that are 2$\sigma$ below the mean S/N for each season. \cite{Wang2020} also use a window of [-2.5 $\times$ MAD, 2.5 $\times$ MAD] to compute $\sigma_{line}$ so as to eliminate the effects of noise and blending in the line wings. 

\subsection{Comparison of the Three Fitting Methods}
\label{Fit_Method_Comparison}
Figures~\ref{Fig:totalMgIILineProfile}, \ref{Fig:totalHbLineProfile}, and \ref{Fig:totalHaLineProfile} show the variability of the \MgII, \Hb, and \Ha\ broad emission-line profiles as quantified by the three different fitting methods.

The three methods result in very similar flux light curves for the \Ha\ and \MgII lines. The different methods also find very similar FWHM measurements for the three lines, implying that the FWHM measurement is robust to the details of the emission-line models \citep{DallaBonta2020}. On the other hand, the three methods have large differences in the $\sigma$ measurement. The $\sigma$ line width is highly sensitive to the wings of the line and so is susceptible to small differences in the fitted line profile. Among the three lines, \Hb\ has the largest differences in measured quantities between the three lines, owing to its complex and asymmetric line profile (shown in Figure \ref{Fig:GIFexample}) that is highly sensitive to small differences in fitting.

Despite differences in the details of the line fits, the three methods result in similar \textit{qualitative} trends for all three emission lines. That is, the differences between methods are generally systematic offsets: e.g., \texttt{QSOFit} measures broader $\sigma$ and lower $\Delta v$ for all lines, as well as brighter \Hb, than the other methods. The similar overall qualitative trends mean that our general conclusions about the relative changes in the line profiles are not dependent on the details of the fitting method.

% In all three emission-line regions, we see qualitatively the same general trend for the line-center velocity shifts between all 3 schemes. There does seem to be a trend that the non-parametric measurements measure the highest line-center velocity shifts between all 3 schemes, the single Gaussian fitting scheme measures the second highest line-center velocity shifts between all 3 schemes, and the \texttt{PyQSOFit} fitting scheme measures the least line-center velocity shifts between all 3 schemes. 

Figure~\ref{Fig:LineBreathingAll} shows the line breathing behavior measured by the three fitting methods. We fit the relationships in each panel using the Bayesian linear regression package \texttt{linmix} \citep{Kelly2007}. The slopes are shown at the bottom left of each panel where $\beta_{QF}$ corresponds to the slope for the QSOFit fitting method, $\beta_{SG}$ corresponds to the slope for the single Gaussian fitting method, and $\beta_{NP}$ corresponds to the slope for our non-parametric measurements. The slopes for FWHM are broadly similar for all three fits to each emission line, while the slopes for line-width $\sigma$ show large differences.
%As described in this section for Figures~\ref{Fig:totalMgIILineProfile}, \ref{Fig:totalHbLineProfile}, and \ref{Fig:totalHaLineProfile}, we observe in the line-profile plots that all schemes produce the same values for FWHM over time while there are discrepancies between the line-width $\sigma$ values between all schemes which is why we see similar slopes here for FWHM and discrepancies in slopes for line-width $\sigma$.
As noted above in the discussion of Figures~\ref{Fig:totalMgIILineProfile}, \ref{Fig:totalHbLineProfile}, and \ref{Fig:totalHaLineProfile}, $\sigma$ line-width measurements are more sensitive to the details of the line profile and so are more dependent on the differences in best-fit models from each method. From this investigation we note that characterizing quasar line breathing with $\sigma$ is likely to depend significantly on the details of the fitting method especially for emission lines that change shape like \target\ (see Figure \ref{Fig:line_width_ratios}). FWHM line-width measurements, on the other hand, are less sensitive to details of the model fitting.

% \subsection{GIF}

% \begin{figure}
% \begin{interactive}{animation}{Animations/Hb.mp4}
 % \includegraphics[width=\textwidth]{exampleFromGIF.pdf}
% \end{interactive}
% \caption{Descriptive caption of example figure and what the animation shows.}
% \end{figure}

% \clearpage
\vspace{1.5in}
\bibliography{lib}{}

\end{document}